\documentclass[aip,pop,amsmath,amssymb,reprint,floatfix]{revtex4-2}

\usepackage[utf8]{inputenc}
\usepackage{graphicx}
\usepackage{bm}

\graphicspath{{figs/}}

\DeclareUnicodeCharacter{00B1}{\ensuremath{\pm}}
\DeclareUnicodeCharacter{03B1}{\ensuremath{\alpha}}
\DeclareUnicodeCharacter{03C0}{\ensuremath{\pi}}
\DeclareUnicodeCharacter{2261}{\ensuremath{\equiv}}
\DeclareUnicodeCharacter{00B5}{\ensuremath{\mu}}
\DeclareUnicodeCharacter{00F3}{\'o}
\DeclareUnicodeCharacter{00F1}{\~n}
\begin{document}
\title{An Enhanced RPA-LDA Model for Ion Stopping Power from Cold Matter to High-Energy Density Plasmas: A Unified, Open-Source Framework}
\author{T.A. Mehlhorn}
\affiliation{Mehlhorn Engineering Consulting Services}
\author{M.F. Gu}
\affiliation{Prism Computational Sciences}
\author{I. Golovkin}
\affiliation{Prism Computational Sciences}
%

\date{\today}

\begin{abstract}
We present an enhanced random-phase-approximation--local-density-approximation
(e-RPA-LDA) model for the stopping power of ions that is valid over a wide range
of conditions, from cold solids through warm dense matter to high-energy-density
plasmas. The electronic stopping is computed from the RPA dielectric response in
the local-density approximation over an average-atom electron density obtained in
a muffin-tin potential with the Flexible Atomic Code, augmented by four
corrections to the earlier RPA-LDA model of Wang et al.: a strong-collision
correction for large-momentum-transfer events, a static local-field correction
for electron correlations, an electron-binding correction, and the higher-order
Barkas and Bloch terms.
The resulting proton stopping powers agree with the NIST PSTAR and IAEA databases
across the periodic table and for compounds---providing a physics-based
alternative to semi-empirical codes such as SRIM---and reproduce the limited
published plasma data, including charged-particle transport-workshop benchmarks,
time-dependent DFT calculations, and the first measurements of enhanced
light-ion stopping in plasmas. We further extend the model to a complete total
stopping power for protons and alpha particles by adding nuclear and ionic
(elastic ion--ion) stopping to the electronic term, yielding a continuous,
self-consistent description of energy deposition from cold matter to hot dense
plasmas. Because the average-atom treatment includes contributions from all
electrons---unlike Kohn--Sham DFT---while remaining computationally efficient and
applicable to low- and high-Z targets at arbitrary temperature and degeneracy,
the model is well suited to inertial fusion and high-energy-density science. The
computational framework is openly available on GitHub
(\url{https://github.com/dedx-erpa/dedx}), with tabulated stopping powers and
ranges for common materials in the \texttt{data/} subdirectory.
\end{abstract}

\maketitle

\section{Introduction}\label{introduction}

The electronic stopping power of energetic ions is of fundamental importance to a broad range of applications in material science, semiconductor manufacturing, medicine, radiation protection, and in fundamental and applied physics. Specific applications include ion beam analysis, ion beam implantation, ion beam machining, ion beam radiotherapy, radiation shielding, and the understanding of ion stopping powers and ranges for a wide array of energetic ions in materials spanning the entire periodic table, including solids, liquids, and gases of both pure and compound materials. The two main databases for electronic stopping power are managed by the Physical Measurement Laboratory of the National Institute of Standards and Technology (NIST) \cite{Berger6130} and the International Atomic Energy Agency (IAEA) \cite{Montari5969}. A recent article reviewed the electronic stopping power within the IAEA database and assessed the abundance and scarcity of available data as a function of energy and collisional systems \cite{Montari5969}. This article also describes how semi-empirical methods, such as SRIM-2013 \cite{Ziegler6266}, are used in the analysis of experimental data. NIST's PSTAR database provides calculated stopping powers, ranges, and related quantities using the underlying methods developed by members of the report committee sponsored by the International Commission on Radiation Units and Measurements (ICRU) \cite{ICRU6267}. The basis for the semi-empirical methods underlying both databases can be found in Anderson and Ziegler's 1977 book on the stopping and range of protons in matter \cite{Anderson4330} which separates the ion stopping power into low- and high-energy contributions, joined by an intermediate interpolation at the stopping power peak. NIST only provides data for proton and alpha particles, while the IAEA database contains datasets for a wider set of heavy ions, although the largest amount of data comes from proton and alpha beam experiments.

The accuracy of the stopping power and ranges provided via these semi-empirical methods are strongly dependent on the quality and selection of the underlying experimental data. Despite the sparsity of the stopping power data and the partial range energy coverage, the "fact that the SRIM-2013 code offers stopping power values for any ion-target system, even for compounds not measured yet by means of Bragg's rule" encourages the sometimes "overconfident" use of the results \cite{Montari5969}. Similar semi-empirical electronic stopping power models are used in Monte-Carlo method radiation transport codes such as Geant and MCNPX, where any deficiencies may be masked in the complicated shower of energetic particles and radiation. Ref \cite{Montari5969} notes that the decrease in the number of publications, measured datasets, and data points produced for cold materials since 2015 can be attributed to the perception that the SRIM-2013 results are good enough and that there are more important experimental measurements to perform. The challenges of measuring the electronic stopping power of low energy ions due to sensitivities to contaminants, inhomogeneities in thickness, and crystallinity are also impediments to attempting to augment the database. We call attention to the concern that the use of SRIM-2013 in the analysis of experimental stopping power data to determine target thickness, inhomogeneities, and contaminant levels could result in the propagation of systematic errors. In this paper, we argue that there is a need for more high-quality experimental data for normal matter to clarify conflicts in existing data and to extend the projectile-target coverage across a broader energy range. We also see the benefit of developing a physics-based, wide range model for electronic stopping power for normal matter, which is the intent of this research.

Ion stopping power is also a fundamental transport property in plasma physics and plays a central role in a wide variety of high energy density laboratory plasmas (HEDLP) experiments, especially charged particle energy deposition in fusion burn waves \cite{Fraley1287,Zylstra6268}. The stopping power of ions has been extensively studied as a primary source of energy deposition in light ion beam fusion and remains of interest in heavy ion beam experiments to generate warm dense matter (WDM) and is the subject of renewed interest in heavy ion fusion (HIF) \cite{Ma5380}. The topic is also especially relevant given the exponential growth in the use of ultrashort pulse laser (USPL) systems to accelerate beams of light and heavy ions to increasingly higher energies and intensities. The stopping power of the ions is central to the generation of many diagnostic signals from inflight atomic and/or nuclear reactions. Recently, there has been significant interest in calculating the stopping power of warm dense matter at the Bragg peak and performing experiments that create and analyze well-controlled WDM conditions \cite{Casas8,Malko9}. Accurate ion stopping power models are essential in both kinetic simulations (e.g. Fokker-Planck collision operator) and as the "ion drag term" in multi-fluid plasma simulations. To advance our understanding of HEDLP science, it is vital to have high-fidelity stopping power and inflight reaction tools that are well-tested across relevant parameter spaces and that are readily accessible to researchers in the HEDLP community. The stopping power of the ions plays a central role in both fast proton ignition and hot spot ignition for inertial fusion energy (IFE) \cite{Ma5380}. For example, the effects of alpha-stopping on ignition and ignition criteria \cite{Reichelt5864} were recent topics of discussion as researchers sought to understand the main impediments to ignition in NIF. In the future, it will also be important to have accurate stopping models for high-Z, non-protonic projectiles in plasmas of arbitrary composition, including high-Z plasmas.

The experimental challenges in obtaining high-quality electronic stopping power data for HED plasmas are even greater than those for normal matter. Plasma stopping power experiments share the same difficulties as normal matter experiments, with the additional complications that a plasma state must be generated by some energetic source, that state must be diagnosed and understood, and in many cases, hydrodynamic motion leads to gradients in density, temperature, and charge state \cite{Barriga5439, Young10, Olsen11, atzeni2004,Ren4167}. To date, most of these experiments have been limited to plasmas with electron temperatures between 1 and 100 eV. A noteworthy exception to this limited temperature regime is the experimental validation of low-Z ion stopping formalisms in HED plasmas in the range of 1.4--2.8 keV and $4\times10^{23}$--$8\times10^{23}$~cm$^{-3}$ performed on the Omega laser \cite{frenje2019}. The achievement of ignition in NIF \cite{Abu-6018} is both an indication that the alpha stopping power in dense DT plasmas is adequately modeled and provides a highly diagnosed platform for gaining further insight into the temperature dependence of the alpha stopping power. Several experimental platforms are being developed in the US, Europe, and China to measure ion stopping power in plasmas, so we expect the validation database to expand.

The remainder of this paper is organized as follows. Section~\ref{History} reviews the development of ion-stopping models that leads to the present work, including our earlier light-ion-fusion stopping models and the unified RPA-LDA model of Wang \textit{et al.} Section~\ref{enhancements-to-rpa-model-beyond-the-previous-unified-model} describes the corrections that define the enhanced e-RPA/LDA model. We then validate the model against cold-matter data (Sec.~\ref{cold-validation}) and warm-dense-matter data (Sec.~\ref{wdm-validation}), add the nuclear and ionic contributions that complete the total stopping power (Sec.~\ref{nuclear-ionic}), and examine energy deposition and the practical sufficiency of the average-atom model (Sec.~\ref{hedlp-deposition}). Section~\ref{future-experiments} proposes experiments at the stopping-power maximum. We then summarize our results and discuss our vision for future work, both in model development and in expanding the experimental database.

\section{Brief Review of Ion Stopping Power Research}\label{History}
The development of theoretical models for charged particle transport in plasmas has been central to the field since the beginning of plasma physics and is essential to all approaches to fusion energy. It is beyond the scope of this article to provide a comprehensive historical review of this topic, but it is important to note that ion stopping power continues to be a topic of significant theoretical interest. The proceedings of two recent workshops on charged-particle transport coefficient comparisons \cite{Grab4212,Stanek13} provide an excellent overview of recent work on a variety of approaches to modeling plasma stopping power, including those based on average atom models, as well as a number of models based on different formulations of density functional theory (DFT) and molecular dynamics (MD). Given that most of this recent work has aimed at understanding ion stopping in fusion plasmas, the theory, modeling, and experiments have been focused on protons and alpha particles stopping in hydrogenic or hydrocarbon plasmas. This trend has also been influenced by the relative ease of accelerating protons using laser-driven sources, compared to heavier ion species.

The electronic stopping power model reported in this paper is a continuation of work that we began in the 1980s as part of the Sandia National Laboratories light ion fusion program \cite{Mehl459}. The goal of that program was to use beams of light ions (protons, lithium, or carbon) that were efficiently generated by modular pulsed power accelerators to implode and burn an ICF pellet \cite{Quintenz1514}, as an alternative to laser fusion drivers. To enable the design of the ICF target, we developed a finite temperature model for the deposition of ion energy by an arbitrary ion traversing a material of arbitrary composition, density and temperature \cite{Mehlhorn447}. The model was then paired with a raytracing algorithm and implemented in the 1-D Titan and 2-D Lasnex \cite{Mehlhorn1147} radiation hydrodynamics codes for use in the ICF target design. The main approach to light ion fusion was termed the "greenhouse target", which was an indirect radiation drive target where the ions penetrated a high-Z (gold) radiation case and deposited most of their energy in a low-density foam, generating X-rays and forming a spherical hohlraum (cf.\ Fig.~5 of Ref.~\cite{Mehl459}). The X-rays within the hohlraum then ablated the shell of an embedded fusion capsule and imploded the DT fuel to thermonuclear conditions. This target design required stopping power models for gold and hydrocarbon foams, $K_\alpha$ imaging \cite{Maenchen3515} and spectroscopy \cite{Bailey4030} of protons incident on thin foils of aluminum or titanium to diagnose the uniformity, transport and focused intensity of the ion beam. Gold foils were also used to Rutherford scatter ions focused to the diode centerline into magnetic spectrometers and Thomson parabolas. This research in ion beam generation and focus produced a 5-MeV proton beam with a power of 9 TW and a microdivergence of 16 mrad that was focused to an intensity of $5.4\pm0.4$~TW/cm$^2$ and heated the first foam-filled hohlraums driven by ions to $35\pm6$~eV \cite{Chandler1517}. In 1993, we focused a 9 MeV lithium beam with a power of 5.5 TW and a microdivergence of 24 mrad to $1.8\pm0.4$~TW/cm$^2$ \cite{Mehlh1518} and reached a hohlraum temperature of $63\pm2$~eV \cite{Derzon457}. $K_\alpha$ spectra from aluminum targets irradiated by the PBFA-II proton beam showed satellite lines up to Al IX (Al$^{+8}$) \cite{MacFarlane353}. These parameters motivate why a wide-range stopping power model was required for low-, mid-, and high-Z targets that accounted for contributions from both bound and free electrons. We deem that the inclusion of this brief history is important because these experiments gave the first ion stopping power in warm dense matter (WDM) regime, but they have been overlooked in recent publications. The results of these experiments, and their position in phase space will be discussed in later sections.

The energy deposition model described in Ref.~\cite{Mehlhorn447} partitioned the total ion stopping into an electronic component for both bound and free (plasma) electrons, and an ion component for both nuclear scattering and plasma ion stopping. The bound electron stopping followed a methodology that is similar to the semi-empirical model found in SRIM-2013, in that it followed a Bethe-like stopping formula for high energy electrons and a Lindhard-Scharff-Schiott (LSS) model for low-energy ions, with semi-analytic estimates for how average ionization potentials and other quantities scaled with the degree of ionization of the plasma. For standalone calculations, a Saha model was used to calculate the plasma $\bar{Z}$ for a given density and temperature. The stopping power of both plasma electrons and ions followed a traditional formulation assuming Maxwell--Boltzmann statistics. A nuclear stopping power term was added to the bound electron stopping power. The model also included a semi-empirical model for the effective charge of the projectile ion that allowed the calculation of the stopping power of high-Z ions. The calculated results were in good agreement with tabulated ranges and stopping powers for cold materials for both protons and uranium. The model demonstrated an initial range shortening with increasing plasma temperature as the target material ionized and the free-electron contribution to the stopping power led to a net increase in the total stopping power. As the temperature continued to increase and the free electron thermal velocity became greater than the projectile velocity, the stopping power decreased, and the range lengthened. (cf.\ Fig.~15 of Ref.~\cite{Mehlhorn447}). The model was partially validated by comparing with data from the first enhanced stopping power experiments performed at the Naval Research Laboratory (NRL) using intense deuteron beams that irradiated Mylar and aluminum targets \cite{Young10}. The results also compared well with proton stopping power experiments performed on the PROTO I accelerator on Mylar, aluminum, and nickel \cite{Olsen11}. We had concerns about the simple analytic scaling of the average ionization potential and related atomic physics in the bound electron model with increasing Z and began looking at more sophisticated atomic physics options, such as the Generalized Oscillator Strength (GOS) model \cite{McGuire1148}. At this time, we also began to conceive of using Lindhard's dielectric response method combined with the local density approximation (LDA) to calculate these terms. We will return to these initial stopping power experiments in Section~\ref{VB}, where we will compare the model presented in this paper with the data and previously published calculations.

\subsection{Unified RPA ion stopping power model}\label{Unified RPA ion stopping power model}
Our next advance in calculating the stopping power of ions in ICF-relevant plasmas was to develop a unified self-consistent model that did not artificially separate the bound from the free electrons and was not divided into low- and high-energy approximations \cite{Wang1}. Rather than using the Lindhard dielectric response method to calculate corrections to terms in the semi-empirical models in the previously discussed stopping power code, this model included sophisticated treatments for the electron density distribution of an atom in plasmas and a full random phase approximation (RPA) stopping function that extrapolated the zero temperature Lindhard stopping function to arbitrary temperatures. This RPA stopping power function was calculated in the local density approximation (LDA) for an electron density distribution function that was calculated in a self-consistent-field "atom in jellium" model \cite{Liberman3892} in a "muffin-tin" configuration. This model has most of the simplicity of an isolated atom model, but captures much of the band structure. This unified RPA-LDA model was in reasonable agreement with the experimental NIST data for both neutral aluminum and gold and was significantly more accurate than the results calculated using an isolated atom model. These calculations relied on the zero temperature Lindhard function. We then implemented Maynard and Deutsch's \cite{Maynard3} temperature-dependent stopping function and compared the finite temperature RPA-LDA results with calculations for the electronic proton stopping power from the scaled Bethe model, as well as some using the GOS model. The RPA-LDA model showed a monotonic decrease in the proton range with ionization up to a $\bar{Z}$ of 20 in a gold plasma. The results fell between those for the scaled-Bethe and GOS models and await future experiments to provide validation data. Although the neutral matter calculations came close to the NIST data for a representative sample of materials across the periodic table, there were sufficient differences in the stopping power peak and low energy regions to not be fully satisfied with the model. Further, the combined Liberman and finite-temperature RPA models were not easy to use, and it was not practical to distribute the code to the plasma physics community. A three-part analytic fit for the stopping power results was generated with the idea that tables could be built to use the results in a radiation-hydrodynamics code, but this was never implemented. Ref.~\cite{Wang1} concluded with the observation that the model was in the framework of the first-Born approximation for the projectile and that the approximation was not appropriate for low-energy heavy ions. The paper ends with the suggestion that higher order Born corrections, including the Barkas and Bloch term be included in the model. Incorporating these higher order corrections is the core advancement in this paper. In the next section, we revisit the unified RPA-LDA model and describe the enhancements that we have made to include higher order Born corrections. We then compare the results of the enhanced RPA/LDA (e-RPA/LDA) with data for stopping in normal (cold) materials and compare the results with experimental data from NIST and IAEA databases. We then compare the results of our e-RPA/LDA model with data and other calculations for plasmas, including the WDM region. 

\section{Enhancements to RPA model beyond the previous unified model}\label{enhancements-to-rpa-model-beyond-the-previous-unified-model}
The main effort of this research has been the development and validation of a comprehensive, state-of-the-art electronic stopping model for ions in HED plasmas based on the unified model of the finite temperature random phase approximation formalism of Wang, Mehlhorn, and MacFarlane \cite{Wang1} by including enhancements to the RPA formalism (e-RPA) beyond the first Born approximation. The e-RPA model is combined with the local density approximation as applied to electron distribution functions (e-RPA/LDA). In this enhanced model we have replaced the Liberman ``atom-in-jellium'' model with electron distribution functions computed using the Flexible Atomic Code, FAC \cite{Gu2}. FAC calculates various atomic radiative and collisional processes, including radiative transition rates, collisional excitation, and ionization by electron impact, energy levels, photoionization, and autoionization, and their inverse processes radiative recombination and dielectronic capture. FAC has been extensively utilized in various research studies for simulating steady-state plasmas under non-local thermodynamic equilibrium conditions, particularly in photoionization-dominated plasmas. A primary goal of the FAC code is to integrate multiple atomic processes within a unified theoretical framework, ensuring consistency across different components and providing a user-friendly interface for accessing computational tasks efficiently. A key factor in using FAC as part of this unified model is that it is a publicly available software package for computing essential atomic data, with a focus on integrating diverse atomic processes cohesively and its demonstration of numerical techniques for efficient computation, thereby benefiting research in astrophysics, plasma physics, and related fields. The resulting enhanced unified e-RPA/LDA model once again avoids the unphysical jumps in stopping power that can occur when bound and free electrons are treated separately. Our goal has been to provide a single comprehensive, efficient, and robust framework for computing energetic ion energy deposition and in-flight reactions in HED plasmas spanning the relevant parameter space. 

\subsection{RPA dielectric response and local density approximation}

In the dielectric response formalism, the electronic stopping power of an ion with charge \(z\) and velocity \(v\) in a uniform electron gas of density \(\rho\) can be written as
\begin{equation}
   \begin{split}
   S_e = -\frac{dE}{dx} = 4\pi\left(\frac{z}{v}\right)^2\rho L(\rho,v) = \\ \frac{2}{\pi}\left(\frac{z}{v}\right)^2\int_0^\infty\frac{dk}{k}\int_0^{kv} d\omega \omega {\rm Im}\left(\frac{1}{\epsilon(k,\omega)}\right)     
   \end{split}  
\end{equation}

where $\epsilon(k,\omega)$ is the dielectric response function, and $L(\rho,v)$ is the stopping function. \\

For electrons with non-uniform density distribution, we apply the local density approximation, and compute the total stopping power as:
\begin{equation}
    - \frac{dE}{dx} = 4\pi\left( \frac{z}{v} \right)^{2}\int d\overrightarrow{r}\rho(\overrightarrow{r})L(\rho,v).
\end{equation} \\
Here $z$ is the charge of the projectile ion. For the light ions considered in this paper it is fully stripped and equal to the nuclear charge, so all results reported here (protons and alpha particles) are free of charge-state ambiguity. For partially stripped heavier projectiles we retain the first-order velocity-dependent effective-charge model of Wang \textit{et al.}~\cite{Wang1}---a Betz-type scaling, $z_{\rm eff}=Z_1[1-1.034\,e^{-v/(v_0 Z_1^{0.69})}]$ in the implementation---which reproduces the $Z_1^2$ dependence of the stopping in cold and warm matter. Because the charge state of a fast ion in a hot plasma can depart substantially from this cold-matter equilibrium---a key requirement of the experiments discussed in Sec.~\ref{future-experiments}---a self-consistent plasma charge-state model remains future work.

In our model, the density distribution surrounding an ion is assumed to be spherical symmetric and calculated in the average atom model (AA) with a muffin-tin-type potential, taking into account the screening effect of the continuum electrons. As mentioned above, the Flexible Atomic Code \cite{Gu2} has been used for the AA calculations. At finite temperature, the dielectric function in the random phase approximation \cite{Zwicknagel4} is used, and therefore we refer to the method as RPA-LDA approximation. The RPA method in our enhanced model uses the same formulation of the Maynard and Deutsch \cite{Maynard3} formulas for the temperature-dependent stopping function, including the interpolation formula developed by Wang et al \cite{Wang1} for L(T,$\rho$,V) that bridges the accurate asymptotic expression for the stopping number in both the small and large projectile velocity limits. The unified RPA/LDA in Ref.~\cite{Wang1} was applicable to cold neutral materials, warm dense matter, and hot plasmas, although the results for neutral atoms agreed better at large velocities (the Bethe regime), than at the stopping power peak and the low energy regime. We now discuss the various correction terms that we have added to the RPA formulation to address these shortcomings. We have implemented the following correction terms to improve the accuracy and range of applicability of the RPA-LDA approximation.

\subsection{Correction terms in the strong coupling regime}

As described by Zwicknagel, Toepffer, and Reinhard \cite{Zwicknagel4}, standard approaches to the energy loss of ions in plasmas when using the dielectric linear response or the binary collision model are strictly valid only in the regimes where the plasma is close to ideal and the coupling between projectile-ion and the plasma target is sufficiently weak. The motivation behind their study of this regime was the application of these methodologies to heavy ion driven inertial fusion and the cooling of beams of charged particles by electrons. They extended conventional linear mean-field treatments using many-body methods and particle simulations to account for strong correlations between the particles and for nonlinear coupling. They reported that quantum effects such as the wave nature of the electrons and Pauli-blocking reduce the stopping power by mollifying the effective interactions. We implemented these corrections by accounting for the fact that the integration over the momentum transfer, k, must be truncated at a certain maximum, \(k_{m}\), above which the collisions become too strong for the perturbation theory underpinning the linear response formalism to be valid. In the regime where \(k > k_{m}\), the contributions are instead calculated with binary collision theory as described by equations~55--59 in Ref.~\cite{Zwicknagel4}.

\subsection{Local field correction terms}

When the electrons in metals, laser-compressed plasmas, and the interiors of heavy planets form a strongly coupled system, RPA is not applicable and the standard Lindhard function needs to be extended. This occurs when the Coulomb coupling constant \(r_{s} \equiv (3/4\pi n)^{1/3}\, m e^{2}/\hbar^{2}\) is greater than unity. Following the polarization-potential approach, strong coupling effects are introduced through local field correction \(G(k,\omega)\), so that the dielectric function is modified to be:
\begin{equation}
  \epsilon_{LFC}(k,\omega) = 1 - \frac{1-\epsilon(k,\omega)}{1+[1-\epsilon(k,\omega)]G(k,\omega)}
\end{equation}    
We have experimented with various formulations of dynamic local field correction \(G(k,\omega)\), including those based on the Mermin dielectric function, as well as some other interpolation methods. We have found the best improvements in accuracy using the static local field correction \(G(k,\omega = 0)\), of Ichimaru and Tago \cite{Ichimaru5} for our models. Therefore, we have chosen to adopt this method rather than the more complex methods associated with dynamic local field corrections. They report that their fitting formula for the dielectric function of a strongly coupled degenerate electron liquid satisfies several self-consistency conditions and accurately reproduces quantum Monte Carlo results. Their static local field correction is given by

\begin{equation}
    \begin{split}
  G(q) = AQ^{4} + BQ^{2} + C + \left\lbrack AQ^{4} + \left( B + \frac{8}{3}A \right)Q^{2} - C \right\rbrack \\
  \times \frac{4 - Q^{2}}{4Q}\ln\left|\frac{2 + Q}{2 - Q}\right|
    \end{split}
\end{equation}
where
\begin{equation}
\begin{split}
 A = 0.029\ \ \ \left( 0 \leq \ r_{s}\  \leq 15 \right) \\
 B = \ \frac{9}{16}\gamma_{0} - \ \frac{3}{64}\left\lbrack 1 - g(0) \right\rbrack - \ \frac{16}{15}A \\
 C = \  - \frac{3}{4}\gamma_{0} + \frac{39}{16}\left\lbrack 1 - g(0) \right\rbrack - \ \frac{16}{15}A
\end{split}
\end{equation}

and r\textsubscript{s}, \(\gamma_{0},\) and g(0) are defined by equations~(4), (5), and (7) of Ref.~\cite{Ichimaru5}. They report that this formula satisfies the self-consistency conditions in the compressibility sum rule and the short-range correlation. They further note that this formulation does not account for dynamic frequency-dependent effects that can be important in describing the detailed features in the spectral function of the density-fluctuation excitations, but we have found it to be adequate for this average atom approach to calculating electronic stopping power.

As noted above, the standard RPA formulation for the dielectric function is only appropriate for ideal targets, and nonideal targets require higher corrections \cite{Ichimaru5}. Interestingly, density functional theories are usually used to incorporate higher correlations into an effective interaction that maintains the simple structure of the linear response. The Kohn--Sham equations with energy-density functionals derived from the local density approximation (LDA) are often used. The alternative to this is to use a local field correction (LFC) technique, such as the one we have implemented. Density functional theory (DFT) is a powerful technique for calculating the electronic structure of atoms, molecules, and solids, especially solid-state band structures. Finite temperature extensions to DFT and molecular dynamics (MD) have also been used to calculate the electrical conductivity for warm plasmas \cite{Dejarlais1142}, Hugoniots and equations of state (EOS) \cite{Desjarlais2583}, as well as other electrical properties, such as the reflectivity of shocked plasmas \cite{Desj2102}. The success of DFT in providing conductivity, reflectivity, and EOS data has encouraged exploration of its application to calculate the electronic stopping of WDM. For example, White et al. \cite{White7} used a mixed stochastic-deterministic time-dependent density functional theory (mDFT) to calculate the proton stopping power in a carbon plasma at $T=10$~eV and various densities. However, the mDFT calculation only treats electrons 2s and 2p in detail, and the 1s contributions are estimated using the model of Casas, Barriga-Carrasco and Rubio \cite{Casas8}. Although we will compare our results with those of mDFT and experimental data in Section \ref{wdm-benchmarks}, it is important to note that our e-RPA/LDA calculations include LFCs and other corrections that incorporate the physics found in the DFT formalism into an average atom approach that is much more computationally efficient. Further, our RPA/LDA code inclusively treats plasma, outer shell, and inner shell electrons within a single unified model, avoiding potential pitfalls that can arise from the combination of results from disparate calculations. Moreover, we have found e-RPA/LDA to be accurate and computationally practical for high-Z elements, such as gold, for which DFT calculations become increasingly challenging. By making our e-RPA/LDA model open source and available on GitHub, we hope to encourage the broader stopping power community to further explore and quantify the tradeoffs between DFT and LFC approaches to effective interaction physics.

\subsection{Binding frequency correction}

We also correct for the fact that the RPA-LDA loss function $L(\tilde{\rho},v)$ depends only on the local density and does not distinguish whether the electrons are bound or free. Our starting point is the discussion by Esbensen and Sigmund \cite{Esbensen6} on the constraints on electron motion caused by local binding forces. Their analysis describes the impact of the simultaneous presence of a single-electron binding frequency and a plasma frequency on distant collisions in a free electron gas. To that end, we introduce an effective electron density $\tilde{\rho}$, such that the loss function at any given radius is given by $L(\tilde{\rho},v)$, and the effective density is defined through a correction to the plasma frequency,

\begin{equation}
    \begin{split}
    \omega_{p}^{2}  =  4\pi\rho \\
    {\widetilde{\omega_{p}}}^{2}  =  4\pi\widetilde{\rho} \\
    {\widetilde{\omega_{p}}}^{2}  =  \omega_{p}^{2} + \gamma^{2}\omega_{b}^{2},
    \end{split}
\end{equation}

where $\omega_{b}$ is the mean binding energy of the electrons. In an average atom mode, and at any given radius, the electron density has contributions from different orbitals, although generally one or two subshells dominate. We calculate the mean binding energy as:
\begin{equation}   
\text{ln}\omega_{b}(r) = \sum_{i}^{}f_{i}(r)\text{ln}I_{i}
\end{equation}

where $f_{i}(r)$ is the fractional contribution of the \(i\)-th bound shell at radius \(r\), and \(I_{i}\) is the binding energy of that shell. In this correction term, the empirical parameter \(\gamma\) is adjusted to match the experimental measurements of the stopping powers in cold materials. We have found that \(\gamma^{2} = 0.75\) provides the best fit to the experimental data and this is the value that we use in our model.

\subsection{Barkas effect}

The Barkas effect, or the effect $Z_1^3$, is the dependence of the stopping power of a heavy charged particle penetrating through a random medium on the projectile charge. The size of the effect was deduced by measuring that the stopping power of antiprotons is $3$--$19\%$ lower than that for protons over the same energy range and is interpreted as a charge dependent polarization effect. Our implementation of the Barkas effect is estimated using the formalism contained in equations 7.12 and 7.13 in Esbensen and Sigmund \cite{Esbensen6}. Expressed in terms of $y = k_{max}v/\alpha_0$ this becomes:
\begin{equation}
    L_{1} = \ \pi\frac{e_{1}e}{mv^{3}}\ \alpha_{0}\ L_{B}(y)
\end{equation}

where
\begin{equation}
    \begin{split}
    L_{B}(y) = \frac{1}{3}(5 - \zeta^2/2)\log y + \frac{1}{3}(1-\zeta^2/4) \\
    \times \left( 1+\frac{y^2}{2}[1-(1+8/y^2)^{1/2}] + 4 log\frac{1+(1+8/y^2)^{1/2}}{4} \right)
    \end{split} 
\end{equation}
with $\zeta^2 = \omega_{0}^2/\alpha_{0}^2$. \\

However, the Barkas term derives from a perturbative expansion ($Z^3$) that diverges at energies below the stopping-power peak, so we suppress it at low projectile energy with an empirical cutoff adjusted to reproduce the cold $dE/dx$ data. In the implementation, the Barkas contribution is multiplied by two factors: a degeneracy prefactor $\propto (v_F^2/v_e^2)^2$ and a projectile energy factor $\exp[-(E_c/E)^{1/2}]$, where $E$ is the projectile energy and $E_c$ is a cutoff energy proportional to the peak stopping-power energy $E_p$ (with a weak, fitted dependence on the target areal density). Here $v_e$ is the electron thermal velocity that includes degeneracy, reducing to the Fermi velocity $v_F$ in the degenerate limit, and $v_F$ is the Fermi velocity. The net effect is to switch the Barkas correction off smoothly once the projectile velocity falls below the characteristic electron velocity; in a hot, weakly degenerate plasma, the $(v_F/v_e)^4$ prefactor renders the Barkas contribution negligible.

At one higher order in the projectile charge we include the Bloch ($Z_1^4$) correction, which bridges the perturbative Bethe and classical Bohr limits and reduces the stopping as the projectile--electron coupling $y = Z_1 e^2/\hbar v$ approaches unity. It is added to the local loss number as $L_{\rm Bloch} = -y^2\sum_{n\ge1}[n(n^2+y^2)]^{-1}$; because $y$ depends only on the projectile charge and velocity, the term is identical in cold matter and in a plasma (a property of the local-density treatment). Like the Barkas term it is a high-velocity expansion that diverges as the stopping number becomes small near the peak, so it is regularized by a smooth velocity cutoff $\exp[-(v_{\rm cut}/v)^4]$ that vanishes at the low-velocity cold-matter peaks and is restored at the higher velocities of hot plasmas, with the single constant $v_{\rm cut}$ fixed so the cold-matter stopping is unchanged.

\section{Validation against cold-matter data}\label{cold-validation}

Verification and validation against the best available data is the organizing principle of this work. We proceed from the most tightly constrained regime -- cold matter, for which large curated databases exist---to the least constrained---high-energy-density plasmas---quantifying the improvement of the enhanced e-RPA/LDA model over the bare RPA at each step. In re-establishing this chain we also draw renewed attention to the groundbreaking light-ion plasma-stopping measurements of the 1980s, which remain the only data in a region of parameter space that modern campaigns have not revisited. We begin with cold matter, where the stopping power is most accurately known and where the empirical parameters of the model are fixed once and for all.

\subsection{Elemental targets}\label{cold-elemental}

We have been developing and implementing strategies to perform rigorous model validation against a variety of data sets. For example, cold stopping data (e.g., NIST) have been used to evaluate the accuracy of the RPA model and explore inclusion of dynamical local field corrections and the Barkas effect in the e-RPA/LDA model.

\begin{figure*}[tp]
\centering
\includegraphics[width=\linewidth]{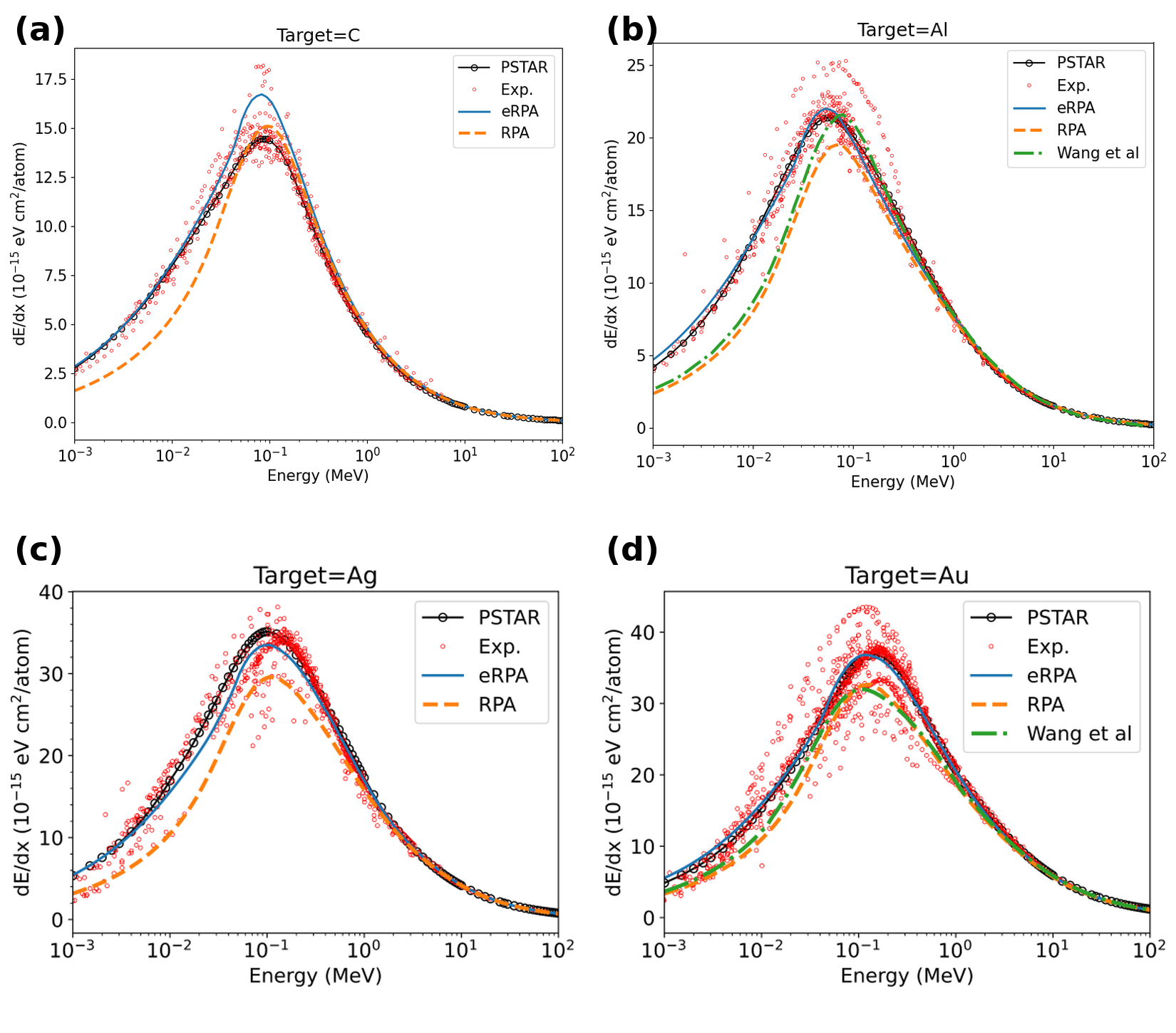}
\caption{Calculated proton stopping power in (a) carbon, (b) aluminum, (c) silver, and (d) gold compared with experimental measurements (red), the PSTAR semi-empirical fit (black), the e-RPA/LDA model (blue, solid), and the uncorrected RPA-LDA (orange, dashed). The corrections bring the model into agreement with the data through the Bragg peak, where the uncorrected RPA-LDA runs low.}
\label{fig:1}
\end{figure*}

As described in the previous section, our model contains some empirical parameters, such as \(\gamma\) in the binding energy correction and the cutoff energy of the Barkas effect. These parameters were adjusted by comparing our computed results with extensive experimental measurements of proton stopping power in cold targets. It is encouraging that the same set of parameters provides a satisfactory fit to measurements over a wide range of atomic numbers. Figure~\ref{fig:1} shows the comparison of our calculated stopping powers of proton in C, Al, Ag, and Au (panels a--d) with the experimental results. These results show that our model provides improved descriptions of the measurements over the original RPA-LDA method, especially at low energies and near the Bragg peak. Quantitatively, taking the semi-empirical values of PSTAR as reference and averaging over the four targets, the e-RPA corrections reduce the mean absolute deviation from about 12\% (RPA) to about 4\% over the full energy range and from about 29\% to about 5\% below the Bragg peak ($E < 50$~keV), where the strong-collision, local-field, binding-energy and Barkas corrections are largest. At the Bragg peak, the corrected model reproduces aluminum, silver, and gold to within 0--5\%, while the uncorrected RPA-LDA underestimates the peak by 10--16\%; carbon is the exception, where the corrected model overshoots the peak by 16\%. The Bragg peak is also where the experimental database is least certain: for gold and silver, the measurements scatter by about $\pm 10\%$ (rms) through the peak --three to four times their $\sim 3\%$ scatter at high energy---so the calculation lies well within the data spread there.

\begin{figure*}[tp]
\centering
\includegraphics[width=\linewidth]{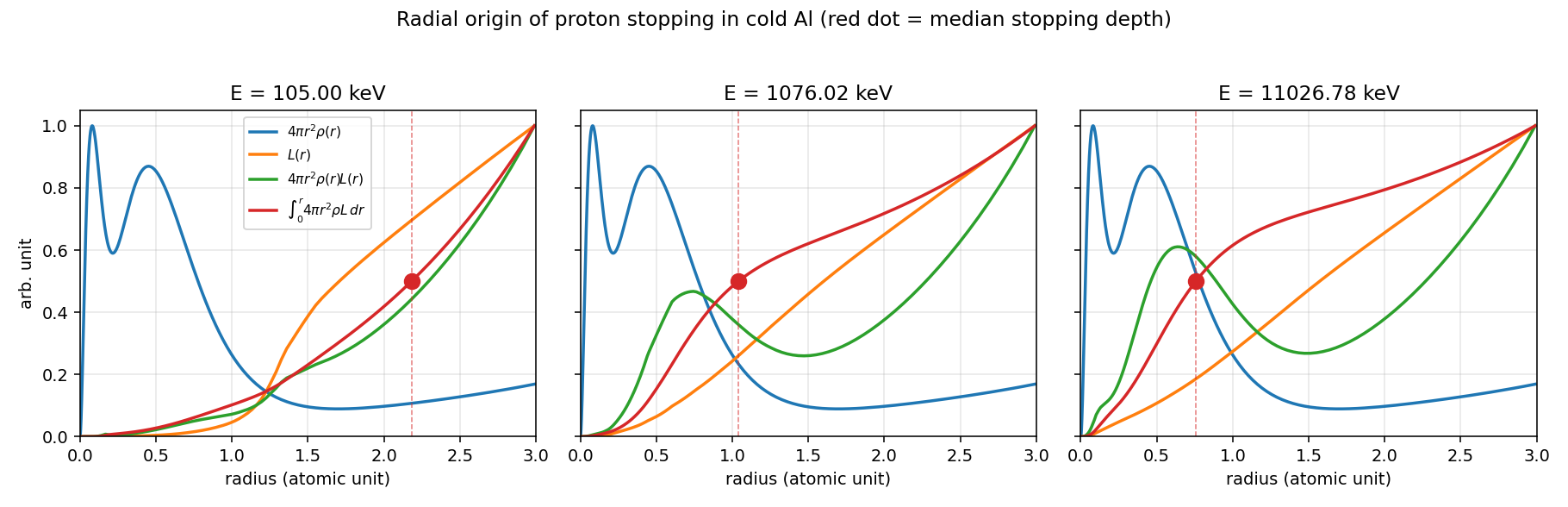}
\caption{Radial origin of the proton stopping in cold Al at proton energies of 0.1, 1, and 11~MeV: the radial electron density $4\pi r^2\rho(r)$, the loss function $L(r)$, the integrand $4\pi r^2\rho(r)L(r)$, and the cumulative fraction of $dE/dx$ within radius $r$. The red marker is the median (50\%) stopping depth, which moves inward as the projectile speeds up.}
\label{fig:5}
\end{figure*}

It is informative to examine the relative contribution at different radii to the total stopping power at various projectile energies. Figure~\ref{fig:5} shows, for a proton in the cold Al target at three energies (0.1, 1 and 11~MeV), the electron density profile $4\pi r^2\rho(r)$, the loss function $L(r)$, the integrand $4\pi r^2\rho(r)L(r)$ of the LDA integral, and the cumulative fraction of the stopping power within a given radius; the red marker indicates the median stopping depth (50\%). As the projectile energy increases, the median stopping depth moves inward (from about 2.2 to 0.8 atomic units), showing that faster projectiles probe the more tightly bound inner-shell electrons.

Finally, the corresponding proton ranges in C, Al, Ag, and Au, obtained by integrating the e-RPA/LDA electronic stopping, track the PSTAR database from high energy down through the Bragg peak; at the lowest energies, the electronic-only range runs above PSTAR because elastic nuclear (ion--ion) stopping, which PSTAR includes and which dominates only in the last few percent of energy, is not yet present. We add that contribution in Sec.~\ref{nuclear-ionic}: the complete proton range (electronic plus nuclear) is shown there in Fig.~\ref{fig:18}, reproducing PSTAR across the full energy range, with panel~(a) isolating the closure of this low-energy overshoot.

\subsection{Compound targets}\label{cold-compounds}

\begin{figure*}[tp]
\centering
\includegraphics[width=\linewidth]{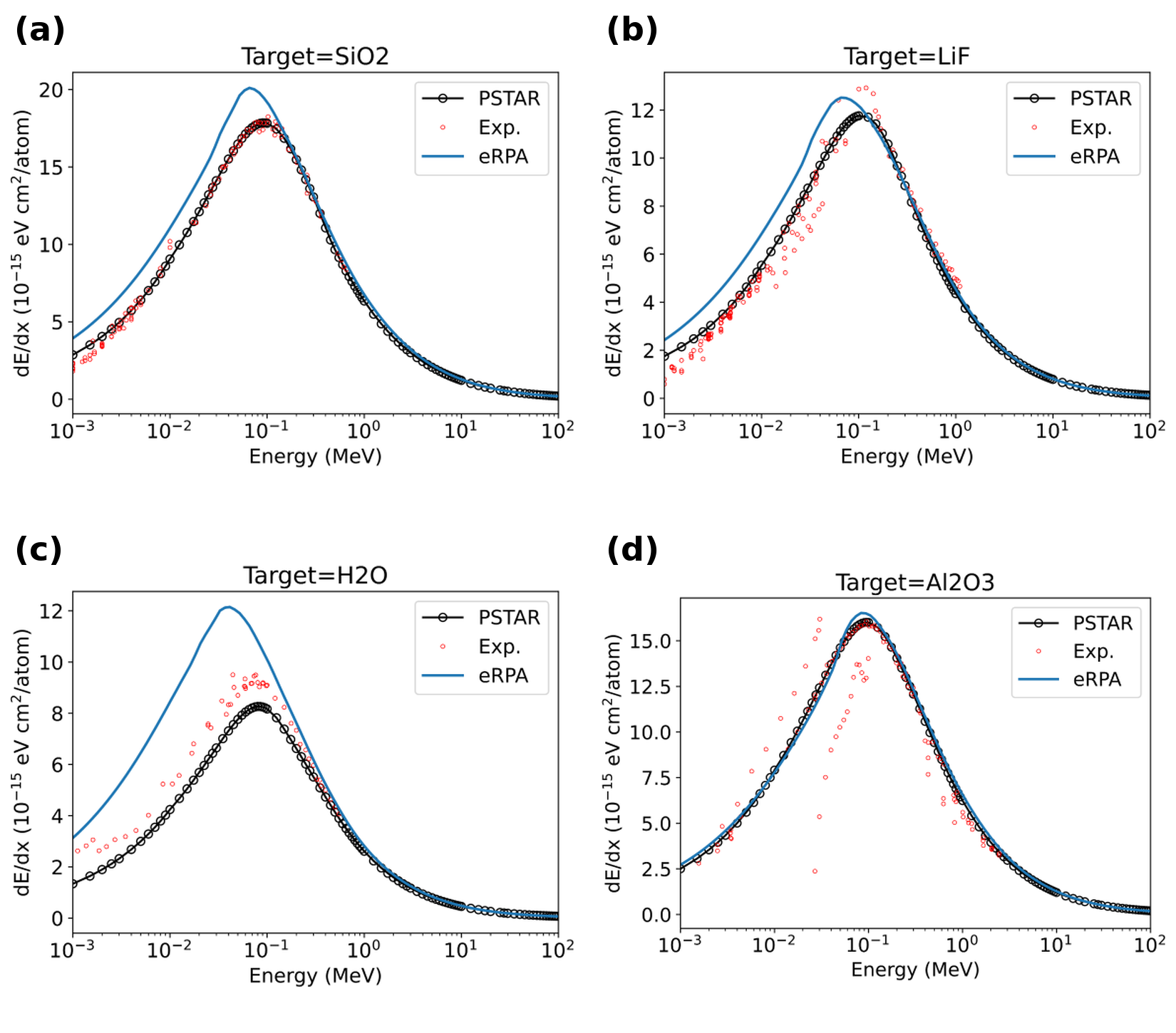}
\caption{Proton stopping power in the compounds (a) SiO\textsubscript{2}, (b) LiF, (c) liquid water, and (d) Al\textsubscript{2}O\textsubscript{3}: average-atom e-RPA/LDA without the band-gap correction (blue, solid) versus the PSTAR fit (black) and experimental data (red). The gapless dielectric response overshoots through the Bragg peak for the wide-gap insulators---most strongly for LiF and water, moderately for SiO\textsubscript{2}, negligibly for Al\textsubscript{2}O\textsubscript{3}---motivating the band-gap correction of Sec.~\ref{band-gap-correction-for-cold-insulators}. Water is anomalous (Bragg additivity breaks down) and shown for completeness only.}
\label{fig:img8}
\end{figure*}

We have developed an average atom model based procedure to obtain the electron density distribution of compounds. Each component atom is assumed to be confined in its own unit cell with radius $r_i$, so that the total volume is equal to that of the unit cell of the compound. The values of $r_i$ are then iteratively adjusted using the average atom model so that the free electron densities at the radius $r_i$ are the same for all constituents. This ensures that all components share the same free electron background and are therefore in pressure equilibrium. The stopping powers are then calculated for each constituent within its own unit cell, the sum of which provides the stopping power of the compound. Figure~\ref{fig:img8} compares the resulting stopping of protons for $SiO_2$, LiF, liquid water, and $Al_2O_3$ (panels a--d) with the PSTAR fit and experimental data. The procedure reproduces the measurements above the Bragg peak in every case; below and through the peak the gapless dielectric response overshoots for the wide-gap insulators---most strongly for LiF and water, moderately for $SiO_2$, and negligibly for $Al_2O_3$---which motivates the band-gap correction developed in the next subsection.

\subsection{Band-gap correction for cold insulators}\label{band-gap-correction-for-cold-insulators}

\begin{figure}[htbp]
\centering
\includegraphics[width=\linewidth]{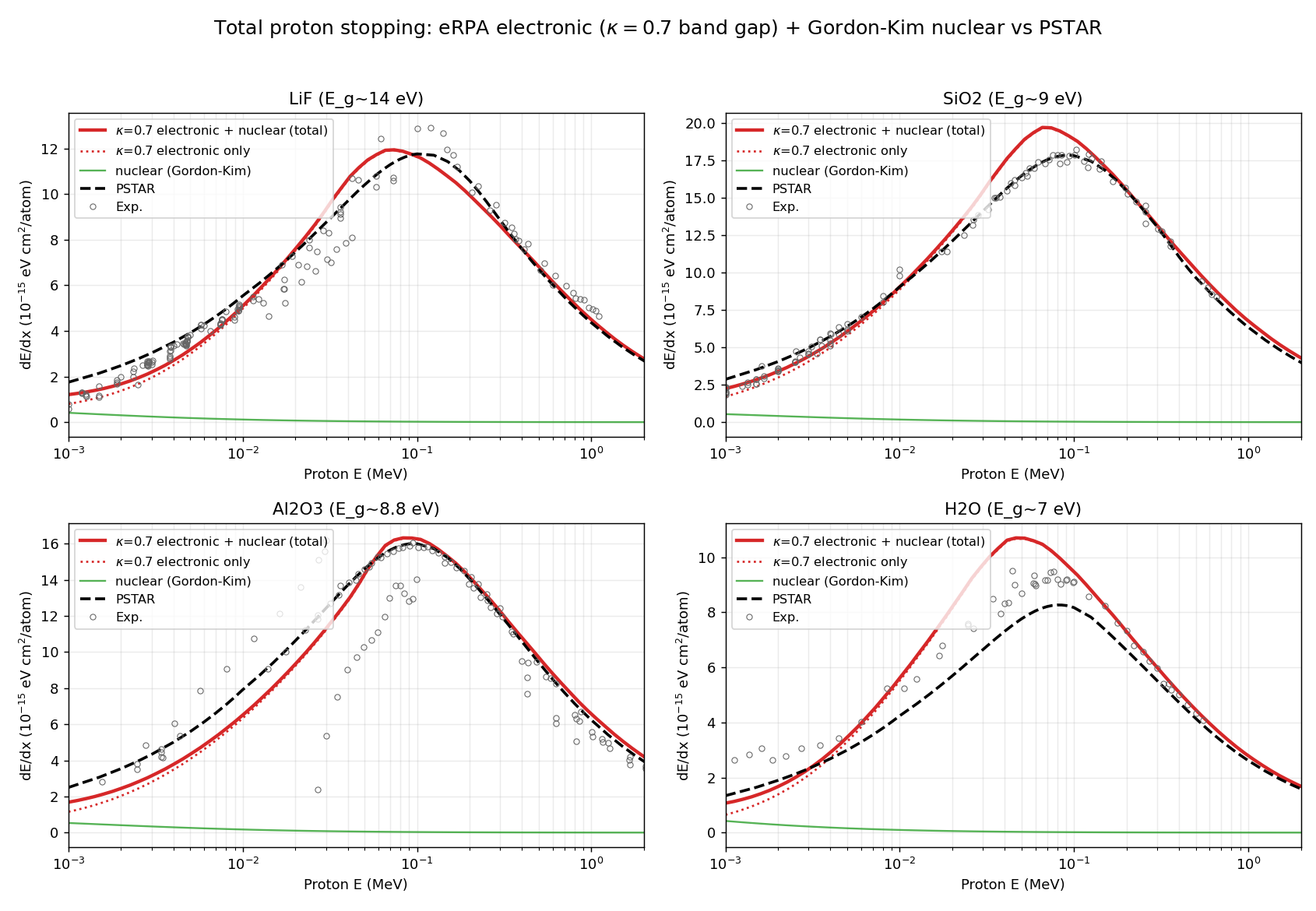}
\caption{Band-gap correction for cold insulators. Total proton stopping ($\kappa = 0.7$ electronic band gap + Gordon--Kim nuclear) for LiF, $SiO_2$, $Al_2O_3$, and $H_2O$ vs PSTAR and experiment; the gapless electronic-only curve (dotted) overshoots through the Bragg peak. Water is anomalous and shown for completeness only.}
\label{fig:27}
\end{figure}

For cold ionic insulators, the average-atom e-RPA/LDA model overshoots the measured electronic stopping through the Bragg peak, because the valence electrons are given a gapless free-electron-gas response that spuriously allows low-frequency excitations forbidden by the material's band gap. We impose a Levine--Louie \cite{levinelouie1982} band gap on the dielectric response: the energy-loss function is zeroed below a threshold $E_g$ and the oscillator strength is shifted upward, preserving the f-sum rule. The threshold is tied to the binding frequency of the atom, $E_g(r) = \kappa\,\omega_b(r)$, where $\omega_b(r)$ is the geometric-mean binding energy of the average-atom~{[}Eq.~(7){]} and $\kappa$ is a single dimensionless constant; because $\omega_b$ derives from the occupations of the average-atom, it evolves continuously with temperature and density, so the correction self-extinguishes as the material ionizes. A single $\kappa = 0.7$ removes the cold overshoot for the wide-gap insulators where the gapless model runs high (Fig.~\ref{fig:27}): the correction scales with the material gap --- largest for LiF ($E_g$ about 14 eV) and smaller for $SiO_2$ and $Al_2O_3$ --- emerging automatically from $\omega_b$ without any per-material tuning. We present $\kappa = 0.7$ as a representative value rather than a precisely determined constant, since the experimental spread between datasets (10--20\%) precludes a tighter calibration. Liquid water, whose stopping is anomalous (breakdown of Bragg additivity, hydrogen-bond-modified dielectric response), lies outside this average-atom, Bragg-additive framework and is shown for completeness only.

\section{Validation in warm dense matter}\label{wdm-validation}

Having fixed the model against cold-matter data, we now apply it in the warm dense matter (WDM) and partially ionized plasma regimes, where experimental data are sparse, and the various theoretical approaches diverge most. We proceed in two steps. We first benchmark the model against the first-principles and average-atom calculations collected in the two charged-particle transport-coefficient comparison workshops~\cite{Grab4212,Stanek13}, against published time-dependent density-functional-theory (TD-DFT) calculations, and against a parameter-free absolute experiment. We then turn to the first measurements of enhanced light-ion stopping in plasmas, performed at NRL and Sandia in the 1980s, which our cold-calibrated model reproduces without further adjustment and which we argue remain an underused constraint on plasma stopping models.

\subsection{Transport-workshop, DFT, and parameter-free benchmarks}\label{wdm-benchmarks}

\begin{figure*}[tp]
\centering
\includegraphics[width=\linewidth]{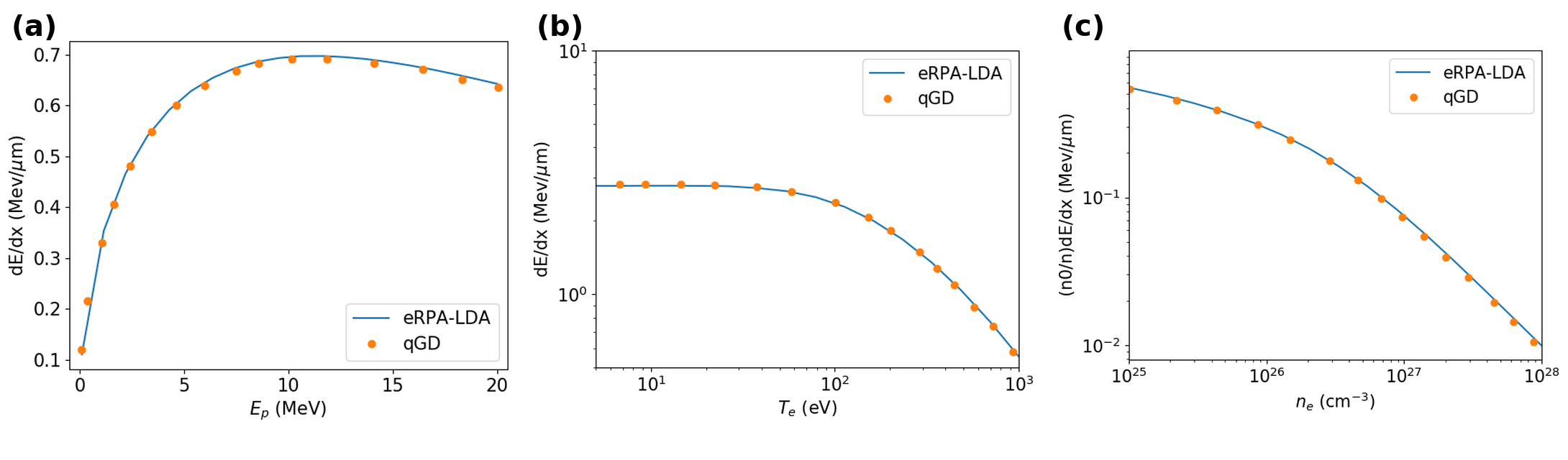}
\caption{Stopping power of an \(\alpha\) particle in a uniform electron gas: e-RPA/LDA (blue) versus the quantum Gould--DeWitt (qGD) benchmark (orange). (a) Versus projectile energy at \(n_{e} = 10^{25}\)~cm\(^{-3}\) and \(T = 1\)~keV; (b) versus electron temperature at a fixed projectile energy of 3.5~MeV; (c) versus electron density at 3.5~MeV.}
\label{fig:8}
\end{figure*}

After calibrating our model using experimental results on cold targets, we applied it to warm dense plasma conditions. The two charged-particle transport-coefficient comparison workshops~\cite{Grab4212,Stanek13} assembled blind comparisons of stopping-power models --spanning average-atom, molecular-dynamics, and several density-functional formulations---against common benchmark conditions and provide the most direct measure of where a given model sits within the modern theoretical spread. We first compare against the quantum Gould--DeWitt (qGD) reference from the first workshop, in which the stopping power of an \(\alpha\) particle in a uniform electron gas was presented; qGD combines a Lenard--Balescu treatment of weak collisions with a T-matrix treatment of close collisions and is a standard benchmark in these comparisons. Figure~\ref{fig:8} compares our model with the qGD reference: panel~(a) versus projectile energy at \(n_{e} = 10^{25}\)~cm\(^{-3}\) and \(T = 1\)~keV, panel~(b) versus electron temperature at a fixed projectile energy of 3.5~MeV, and panel~(c) versus electron density at 3.5~MeV. The e-RPA/LDA model tracks the qGD result across all three.

\begin{figure}[htbp]
\centering
\includegraphics[width=\linewidth]{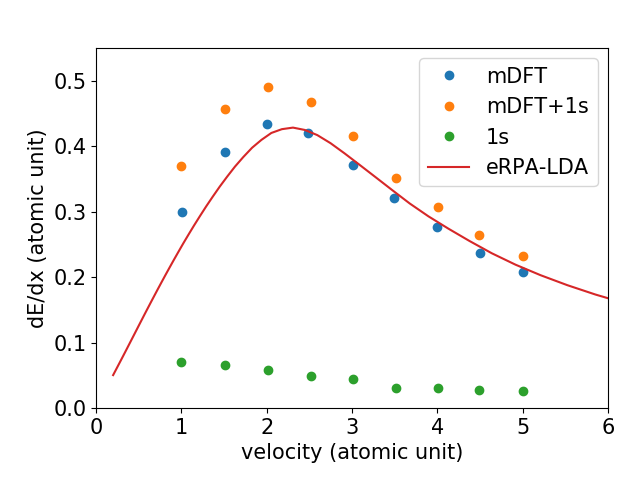}
\caption{Stopping power of proton in a carbon plasma with density 3.5~g/cc and temperature of 10~eV.}
\label{fig:11}
\end{figure}

Next there have been recent developments of applying density functional theory to the calculation of electronic stopping powers. White et al.\cite{White7} used a mixed stochastic-deterministic time-dependent density functional theory (mDFT) to calculate the proton stopping power in a carbon plasma at $T=10$~eV and various densities. Figure~\ref{fig:11} shows a comparison of our model and the results of mDFT for the carbon density of 3.5~g/cc and \(T = 10\)~eV. The mDFT calculation only treats the \(2s\) and \(2p\) electrons in detail, and the 1s contributions are estimated with the model of Casas, Barriga-Carrasco and Rubio \cite{Casas8}. However, it is seen that our model agrees well with the mDFT results without inclusion of the 1s contributions. We estimate that the 1s contributions at energies near or below the Bragg peak should be completely negligible. A radial decomposition of the stopping (as in Fig.~\ref{fig:5}) confirms this: at a proton energy of 33~keV (velocity 1.1~a.u.) the \(1s\) electrons, mostly confined within \(r < 0.5\)~a.u., contribute less than 1\% of the stopping power in the carbon plasma at \(T = 10\)~eV and 3.5~g/cc. We, therefore, find it puzzling that the 1s contributions in Fig. 11 increase with decreasing projectile energy. 

\begin{figure}[htbp]
\centering
\includegraphics[width=\linewidth]{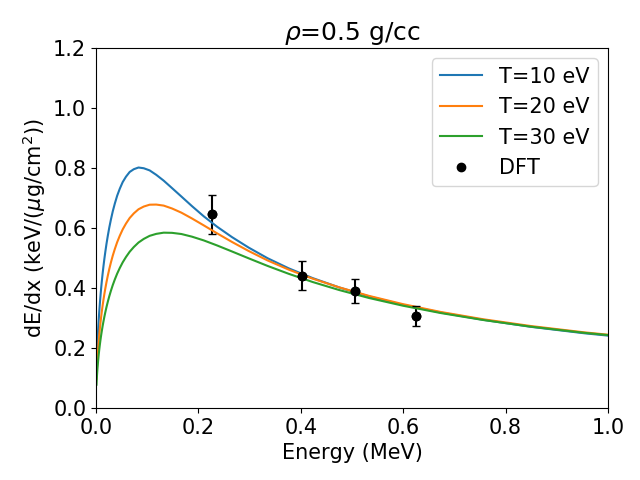}
\caption{Stopping power of proton in a carbon plasma with density 0.5~g/cc and temperatures of 10, 20, and 30 eV.}
\label{fig:13}
\end{figure}

Malko et al. \cite{Malko9} recently published experimental and theoretical results of proton stopping in a warm dense carbon plasma with a density of $0.5$~g/cc, and temperatures of 10, 20 and 30 eV. For the covered proton energies, the results of the density functional theory calculations show little temperature dependence, while several different theories show a wide variation of stopping powers near the Bragg peak. Figure~\ref{fig:13} shows the comparison of DFT and the present model. This Figure shows that our results are in close agreement with DFT calculations for these conditions.

\begin{figure}[htbp]
\centering
\includegraphics[width=\linewidth]{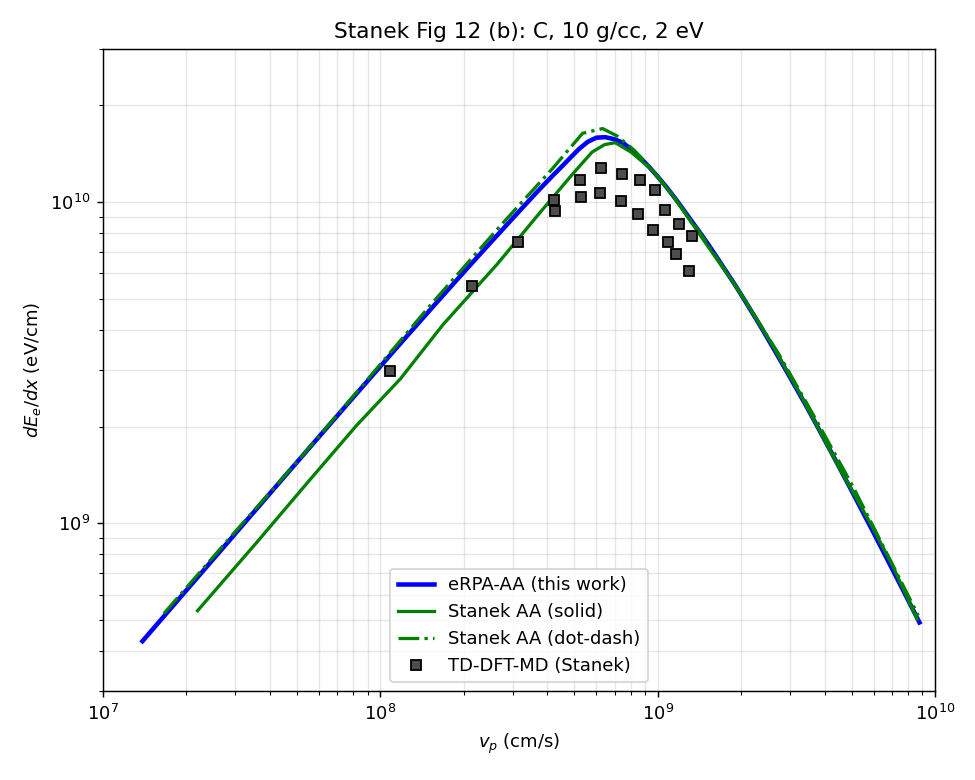}
\caption{Alpha electronic stopping in carbon (10 g/cc, 2 eV) from the Stanek et al. benchmark: e-RPA/LDA (this work) compared with Stanek's two average-atom curves and the TD-DFT-MD reference points, all on the projectile-velocity axis.}
\label{fig:19}
\end{figure}

The same average-atom framework lets us extend the comparison with time-dependent density functional theory begun in Fig.~\ref{fig:11}. The recent multi-code benchmark of Stanek et al. \cite{Stanek13} reports on the TD-DFT-MD reference stopping for alpha particles in hydrogen, carbon, and aluminum, together with the authors' own average-atom results. As in the mixed-DFT calculation of White et al., the TD-DFT-MD treatment calculates the valence electrons explicitly and adds a separately estimated deep-core contribution. Fig.~\ref{fig:19} overlays our e-RPA/LDA result for the carbon case on the Stanek TD-DFT-MD points and on Stanek's two average-atom curves. Our independent e-RPA/LDA peak falls between the two Stanek average-atom curves, and all three average-atom models sit about 20--30\% above the TD-DFT-MD data near the Bragg peak while converging on the common Bethe limit at high velocity. The excess near the peak therefore appears to be a property of the average-atom treatment as a class, not of our particular implementation. This is a good example of where new experimental data could have a significant impact on validating these models. 

The decomposition extended across the projectile velocity (again as in Fig.~\ref{fig:5}) shows that for aluminum the median stopping depth lies among the loosely bound valence and conduction electrons near the Wigner--Seitz radius at low velocity (about 2.4 atomic units) and moves inward to the inner shells only as the projectile becomes fast (to about 1 atomic unit near 1~MeV/u); throughout the velocity range of the Stanek and Malko comparisons the deep core contributes little. We note that the average-atom model captures the inner-shell physics self-consistently over the full velocity range, without the separately estimated core term that the time-dependent density functional calculations must append---a practical advantage of the unified e-RPA/LDA approach.

\begin{figure*}[tp]
\centering
\includegraphics[width=\linewidth]{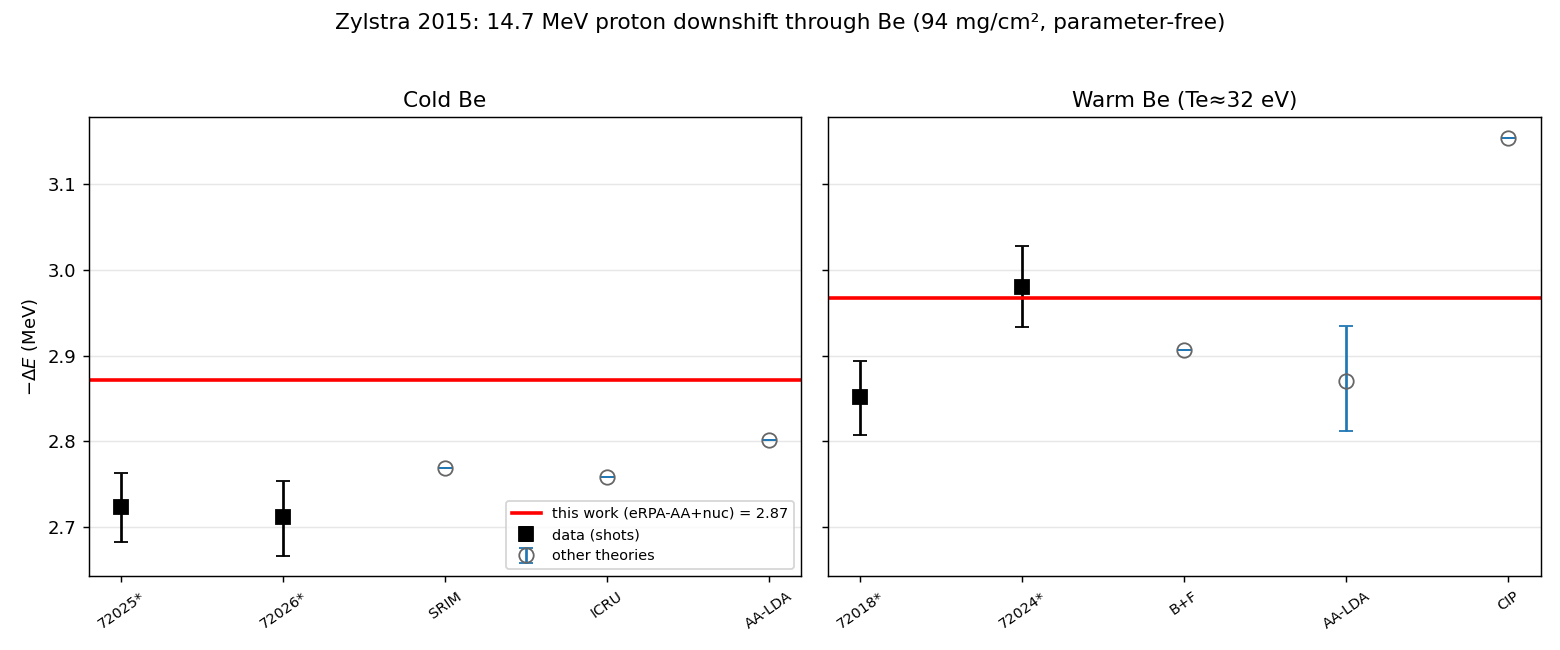}
\caption{Zylstra 2015: energy downshift of 14.7 MeV protons through a 94~mg/cm$^2$ Be foil, cold and warm (parameter-free). This work (red) vs the measured shots and other theories (SRIM, ICRU, AA-LDA, B+F, CIP).}
\label{fig:25}
\end{figure*}

Finally, Zylstra \textit{et al.}~\cite{zylstra2015} provide a parameter-free absolute test: the energy downshift of 14.7 MeV D-$^3$He protons through a $94$~mg/cm$^2$ beryllium foil; both cold and warm ($T_e = 32$~eV). With the areal density taken from the paper and no fitting, our e-RPA/LDA model reproduces the measured downshift --- $\Delta E = 2.87$~MeV (cold) and 2.97 MeV (warm) vs measured 2.72 and 2.85--2.98 MeV --- and matches the authors' own average-atom LDA values (2.80, 2.87), the framework they found consistent with the data, to within about 3\% (Fig.~\ref{fig:25}). Because the foil is flat with a known areal density, this is a genuine parameter-free absolute comparison.

\subsection{The first light-ion plasma measurements}\label{VB}

\begin{figure}[htbp]
\centering
\includegraphics[width=\linewidth]{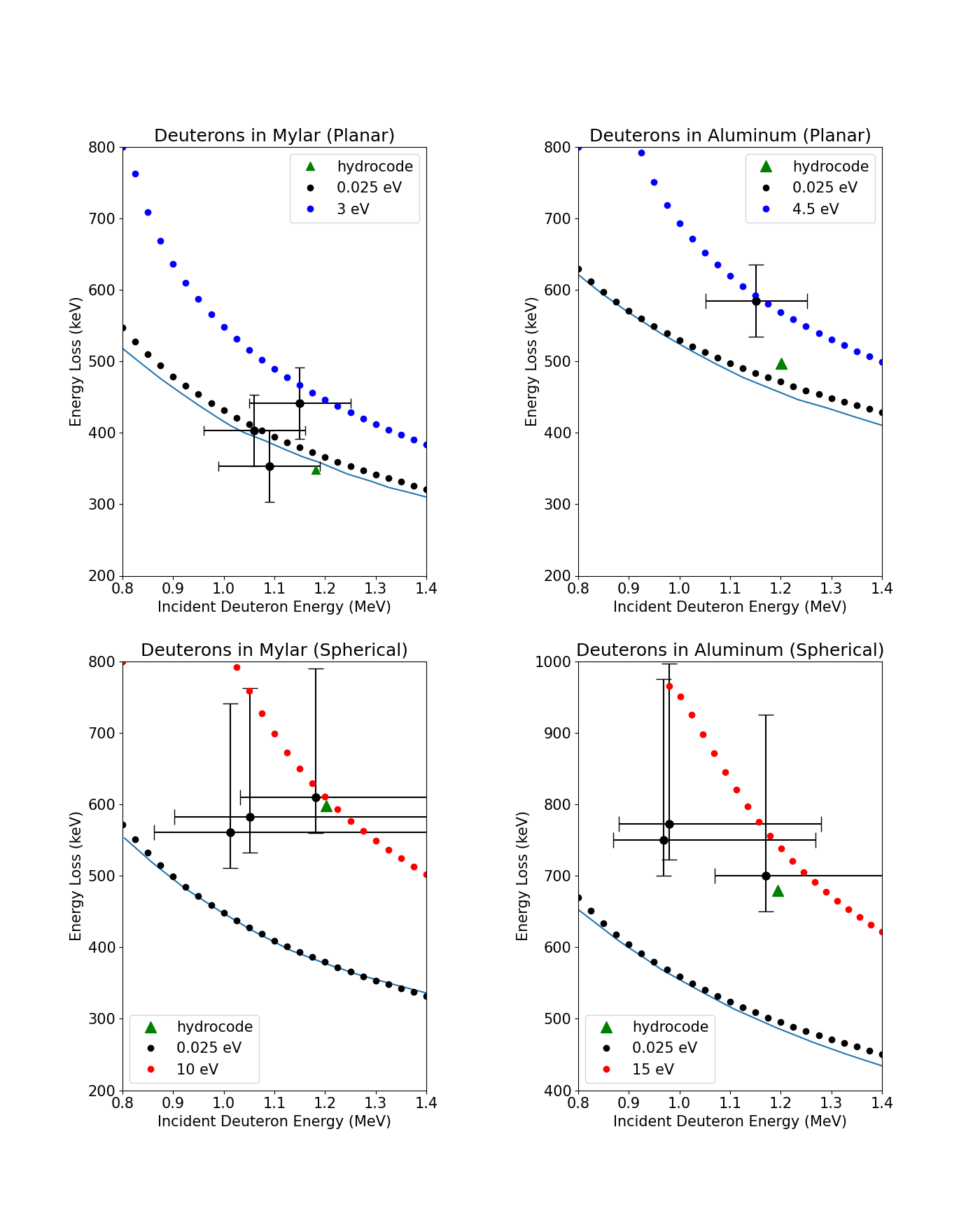}
\caption{Energy loss of 1~MeV deuterons in Mylar and aluminum target-ablation plasmas (after Young et al.~\cite{Young10}, their Fig.~4), for planar (top, $50$~kA/cm$^2$) and spherical (bottom, $250$~kA/cm$^2$) diodes. Points with error bars are the measurements and green triangles the hydrocode results; black points are the cold ($0.025$~eV) e-RPA/LDA calculation (solid line), and the colored points are e-RPA/LDA at the hydrocode-inferred plasma temperatures (Mylar $3$ and $10$~eV; Al $4.5$ and $15$~eV).}
\label{fig:14}
\end{figure}

Young et al.~\cite{Young10} reported the first measurements of enhanced ion stopping in a dense plasma: the energy loss of 1~MeV deuterons in aluminum and Mylar targets heated by an intense light-ion beam, diagnosed with planar diodes ($50$~kA/cm$^2$) and spherical diodes ($250$~kA/cm$^2$). Hydrocode simulations of the experiment give peak electron temperatures of $4$--$5$~eV (Al) and $2.5$--$3.5$~eV (Mylar) at the planar-diode current density, rising to $13$--$17$~eV (Al) and $9$--$11$~eV (Mylar) at the spherical-diode level. Figure~\ref{fig:14} reproduces their Fig.~4 and overlays our e-RPA/LDA energy loss evaluated at both the cold ($0.025$~eV) and the hydrocode-inferred plasma temperatures. The heated e-RPA/LDA calculation reproduces the measured enhancement at the higher current density---about $40\%$ in Al and $45\%$ in Mylar---while the cold calculation matches the lower-current data, supporting the original interpretation that the enhancement arises from stopping by the liberated free electrons in addition to the bound electrons.

\begin{figure}[htbp]
\centering
\includegraphics[width=\linewidth]{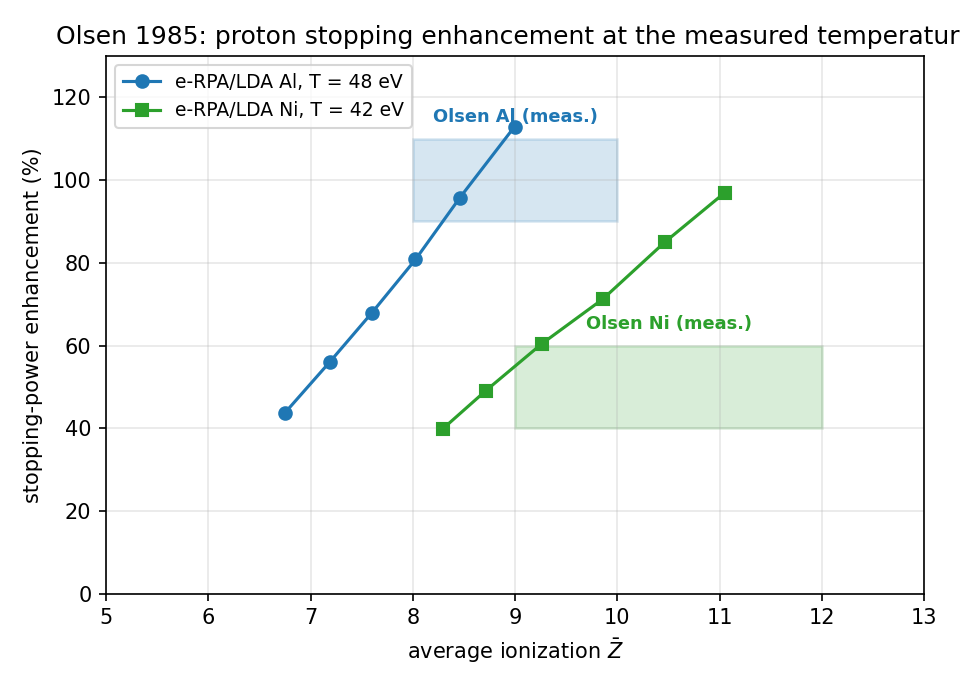}
\caption{e-RPA/LDA stopping-power enhancement for 1.6~MeV protons in Al and Ni versus average ionization $\bar{Z}$, computed as the reduction in the proton range relative to cold matter along the measured peak isotherms ($T=48$~eV for Al, $42$~eV for Ni), with the density fixed by the average-atom ionization balance. Shaded boxes are the enhancements measured by Olsen et al.~\cite{Olsen11} ($\approx 100\%$ for Al at $\bar{Z}=8$--$10$; $\approx 50\%$ for Ni at $\bar{Z}=9$--$12$).}
\label{fig:16}
\end{figure}

Olsen et al.~\cite{Olsen11} measured enhanced proton stopping in beam-heated aluminum and nickel foils, recording the energy loss of a 1.6~MeV proton beam with a time-resolved Thomson-parabola analyzer and interpreting it with a one-dimensional hydrocode simulation of the heated target. At peak intensity, they inferred electron temperatures of about $48$~eV (Al) and $42$~eV (Ni), average ionizations $\bar{Z}\approx 8$ -- $10$ (Al) and $9$ -- $12$ (Ni), and stopping-power enhancements of about $100\%$ and $50\%$, respectively. The enhancement is strongly density dependent, so rather than estimate the expanded foil density we fix the temperature at the measured peak value and obtain the density from the average-atom ionization balance---i.e.\ the density for which the model reproduces the measured $\bar{Z}$. Figure~\ref{fig:16} shows the resulting enhancement of e-RPA / LDA, the reduction in the range of 1.6~MeV proton relative to cold matter, along each measured isotherm. At the consistent density, the model reproduces the aluminum result ($\approx 80$--$110\%$ across $\bar{Z}=8$--$10$) and gives $\approx 60\%$ for nickel at the low-ionization end of the measured range, rising above it at higher $\bar{Z}$---consistent with Olsen's observation that the nickel enhancement is strong and model dependent. The strong density sensitivity (the enhancement more than doubles between $\bar{Z}=7$ and $\bar{Z}=9$ along the aluminum isotherm) underscores that a faithful comparison requires the time-resolved plasma conditions, which is precisely what the hydrocode analysis provides.

\subsection{Summary of WDM verification and validation}\label{VC}
\begin{figure}[htbp]
\centering
\includegraphics[width=\linewidth]{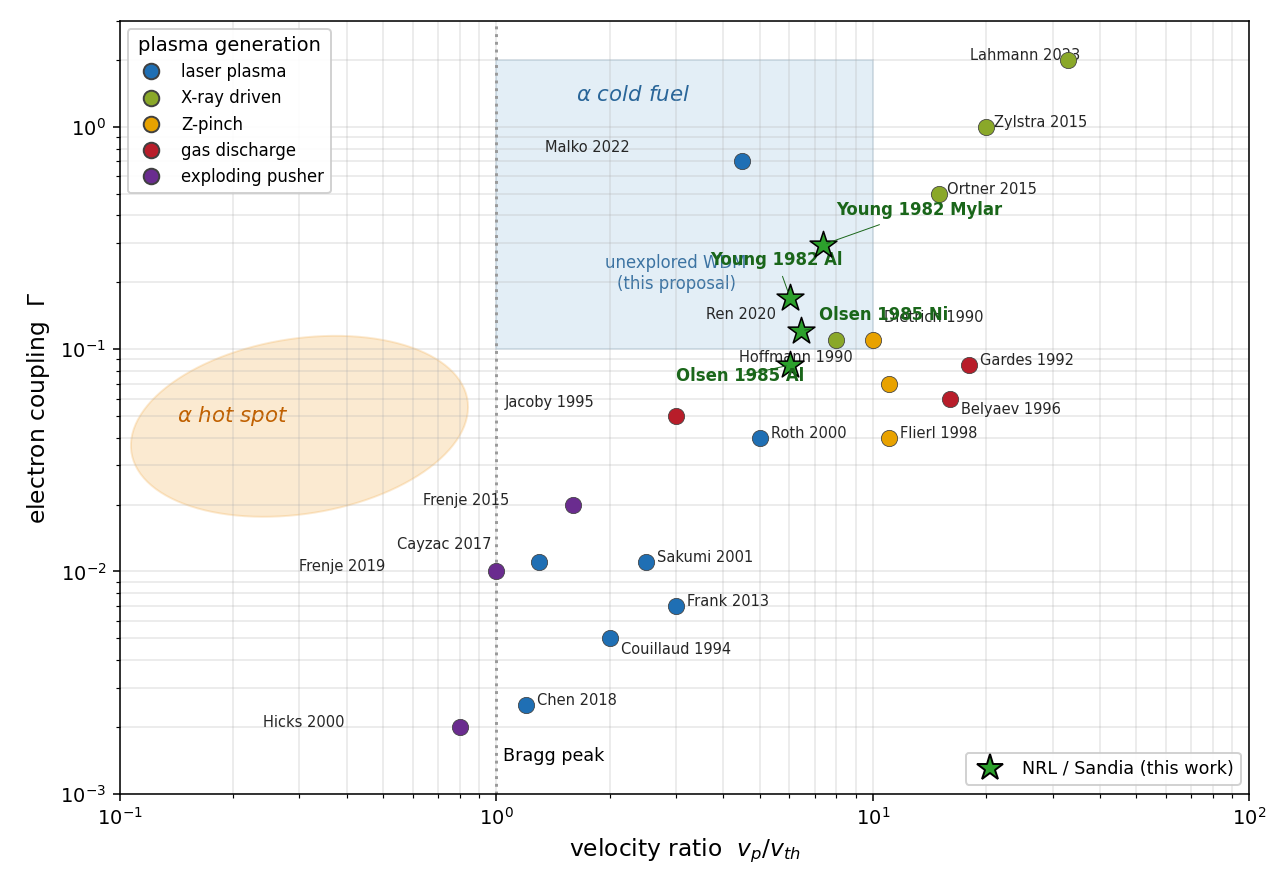}
\caption{Reported ion-stopping experiments in the velocity-ratio ($v_p/v_{th}$) and electron-coupling ($\Gamma$) plane, colored by plasma-generation method (laser plasma, X-ray driven, Z-pinch, gas discharge, exploding pusher). The orange ellipse and light-blue box mark, respectively, the ICF alpha hot-spot and cold-fuel deposition conditions; the cold-fuel box is the warm-dense-matter region (near the Bragg peak, $v_p/v_{th}\lesssim 10$, $\Gamma\approx 0.1$--$10$) where the models diverge most and data are sparsest. The 1980s NRL (Young et al. \cite{Young10}) and Sandia (Olsen et al. \cite{Olsen11}) enhanced-stopping experiments (green stars), evaluated from their published plasma conditions, fall at $v_p/v_{th}\approx 6$--$7$, $\Gamma\approx 0.09$--$0.3$ --- inside the cold-fuel box and among the modern campaigns, yet they remain the only measurements in this regime and have been largely overlooked. The modern-experiment compilation is adapted from Malko et al. \cite{Malko9} and Frenje et al. \cite{frenje2019}; coordinates use $\Gamma = e^2/(a_e k_B T_e)$ with $a_e = (3/4\pi n_e)^{1/3}$.}
\label{fig:paramspace}
\end{figure}

In summary, we find that the e-RPA/LDA model reproduces both the first measurements of enhanced stopping in partially ionized plasmas --- the 1980s NRL and Sandia experiments of Young et al. \cite{Young10} and Olsen et al. \cite{Olsen11} in Fig.~\ref{fig:14} and \ref{fig:16} --- and the modern warm-dense-matter benchmarks of Fig.~\ref{fig:8}--\ref{fig:13}. These early experiments, on mid-Z target-ablation plasmas, occupy a region of the coupling--velocity plane that the recent hydrogenic and hydrocarbon campaigns have not revisited and that the current community has largely overlooked. Evaluated from the published plasma conditions, they lie at a velocity ratio $v_p/v_{th}\approx 6$--$7$ and electron coupling $\Gamma\approx 0.09$--$0.3$ (Fig.~\ref{fig:paramspace}), squarely among the modern campaigns and adjacent to the cold-fuel alpha-stopping condition rather than in an isolated corner of the parameter space. That a single e-RPA/LDA framework, calibrated only against cold-target PSTAR data, matches both these 1980s measurements and the recent TD-DFT and laser-plasma results argues that the older data remain a valuable and underused constraint on plasma stopping models.

The plasma regime of interest is bounded by the electron-coupling and degeneracy parameters
\begin{equation}
\Gamma \equiv \frac{e^2}{a_e\,k_B T_e} \gtrsim 0.1,
\qquad
\Theta \equiv \frac{k_B T_e}{E_F} \lesssim 10,
\end{equation}
where $a_e = (3/4\pi n_e)^{1/3}$ is the radius of the electron-sphere and $E_F$ the Fermi energy.

\section{A complete stopping model: nuclear and ionic contributions}\label{nuclear-ionic}

The validation so far has concerned electronic stopping. To obtain a complete total stopping power---continuous from cold solids to hot dense plasmas, and accurate at the low-velocity end of the range where it dominates---we add the nuclear (elastic ion--ion) contribution. This term plays two distinct roles: at low projectile energy in cold matter it closes the residual range gap left by the electronic-only model, and in hot dense plasmas the projectile-ion/target-ion channel becomes an appreciable, separately measurable contribution to the stopping.

\subsection{Closing the cold-matter range gap}\label{nuclear-coldrange}

\begin{figure*}[tp]
\centering
\includegraphics[width=\linewidth]{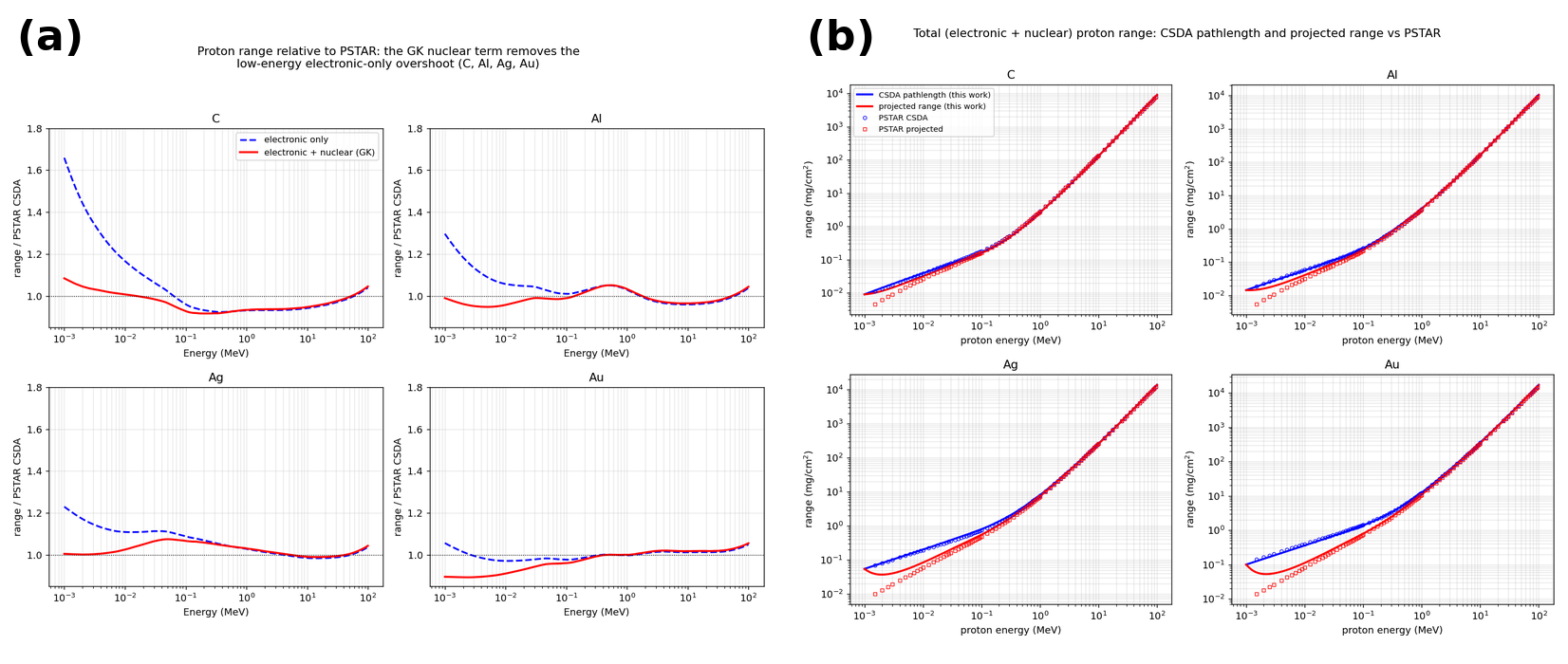}
\caption{Proton range for C, Al, Ag, and Au with and without the nuclear term. (a) Range relative to the PSTAR CSDA range: electronic stopping only (dashed) versus electronic plus Gordon--Kim nuclear (solid); the nuclear term removes the low-energy overshoot of the electronic-only range. (b) Absolute CSDA pathlength and projected (practical) range of the total (electronic + nuclear) model versus the PSTAR CSDA and projected ranges.}
\label{fig:18}
\end{figure*}

We have implemented the nuclear (elastic ion--ion) stopping power following the classical two-body scattering formalism of Faussurier, Blancard, and Gauthier \cite{Faussurier12}, using three screened pair potentials: an ion-sphere potential, a Yukawa potential, and a finite-temperature Gordon--Kim ``peanut-molecule'' potential constructed from the same average-atom electron density used for the electronic term. The Gordon--Kim form, which recovers the bare nuclear product at small separation, is the default. Because the nuclear term is added to the electronic stopping on a common per-target-atom basis, the model now returns a total stopping power that is continuous from cold solids through warm and hot dense matter. Fig.~\ref{fig:18}(a) compares the CSDA range of protons in C, Al, Ag and Au --- computed from electronic stopping alone and from electronic-plus-nuclear total --- against the PSTAR database, whose CSDA range includes nuclear stopping. At the lowest energies, the electronic-only range overshoots PSTAR by 23--66\% for C, Al, and Ag; adding the Gordon--Kim nuclear term brings the range into agreement with PSTAR to within a few percent, closing the low-energy overshoot of the electronic-only range. For gold, the nuclear term slightly overcorrects (about 10\%), consistent with the expected frozen-density over-repulsion of the Gordon--Kim potential at high atomic number.

Our implementation reproduces the Faussurier's original results. For cold proton stopping in aluminum the Gordon--Kim nuclear stopping matches the NIST/ZBL reference to within about $5\%$ (within ZBL's own fit uncertainty), in agreement with Faussurier \textit{et al.}~\cite{Faussurier12}, whose calculations used the proprietary SCAALP average-atom model. Because the Gordon--Kim potential is built from the total electron density -- which integrates with the nuclear charge by the neutral-atom constraint -- the cold nuclear stopping is insensitive to the underlying average-atom model, so the SCAALP and FAC results coincide; the FAC mean ionization ($\bar{Z}=3.0$ cold, rising to about $12.8$ at $1$~keV) likewise matches the SCAALP values. Beyond reproducing the model, our implementation differs in three respects. First, the nuclear term is built from the \emph{same} FAC average-atom density used for electronic stopping, so the electronic and nuclear contributions form a single self-consistent average-atom description rather than two separate calculations. Second, the Gordon--Kim overlap is evaluated with a muffin-tin (embedded-in-medium) boundary consistent with the e-RPA electronic treatment, which softens the nuclear stopping by about $5\%$ in hot dense matter where the interstitial electron density is appreciable. Third, and most consequential for the community, the entire pipeline is built on the open-source Flexible Atomic Code: the nuclear/ionic stopping that previously required the proprietary SCAALP code is now reproducible end to end in a fully open framework.

The same nuclear scattering model also provides the angular straggling that separates the CSDA pathlength from the projected (practical) range tabulated by Ziegler/SRIM, Janni, PSTAR, and the IAEA. The deflection integral evaluated for nuclear stopping, $S = \int \sin^2(\theta/2)\, p\, dp$, is the momentum-transfer (transport) cross section $\sigma_{tr} = 4 \pi S$ that drives the projectile's multiple-angle scattering; propagating it along the track reproduces the PSTAR projected-to-CSDA range ratio for protons in aluminum from 0.1 to 10 MeV. The public package now writes both ranges. The detour is negligible (above 0.98) over most of the range and matters only in the final few percent of energy --- the end-of-range region where the nuclear contribution to stopping becomes significant.

The total (electronic + Gordon--Kim nuclear) model also produces the two ranges that the standard tables --- Ziegler/SRIM, Janni, PSTAR, and the IAEA --- quote: the CSDA pathlength and the projected (practical) range. Fig.~\ref{fig:18}(b) shows both protons in C, Al, Ag, and Au against PSTAR. The CSDA path length now includes the nuclear energy loss, which removes the low-energy range overshoot of the electronic-only model (cf. panel a); the projected range additionally accounts for the nuclear multiple-angle scattering through the transport cross section $\sigma_{tr} = 4\pi S$. Both reproduce the PSTAR values over the full energy range, with the projected range separating from the pathlength below about 0.1 MeV, where end-of-range scattering becomes important.

\subsection{Ion stopping in high-energy-density plasmas}\label{nuclear-hedp}

\begin{figure}[htbp]
\centering
\includegraphics[width=\linewidth]{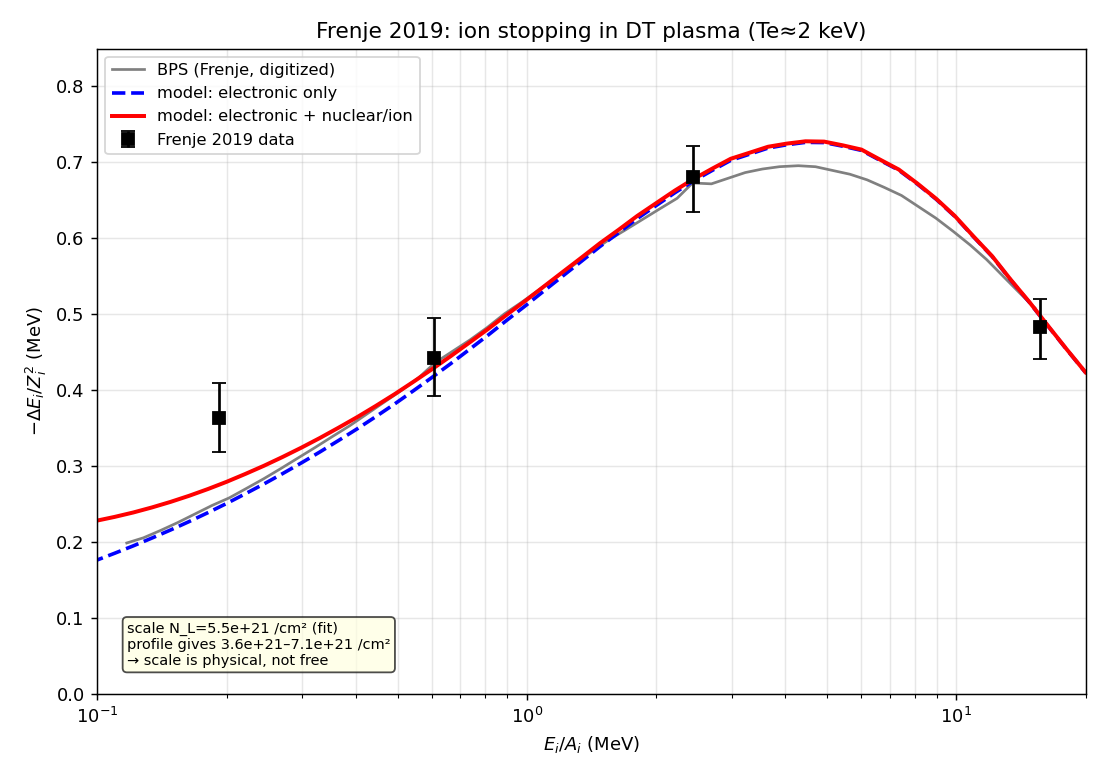}
\caption{Frenje 2019: ion stopping in a DT plasma ($T_e = 2$~keV). Model electronic-only (dashed) and electronic + nuclear/ion (red) vs the digitized data and BPS; the nuclear/ion channel lifts the low-velocity stopping toward the measured excess.}
\label{fig:24}
\end{figure}

Frenje \textit{et al.}~\cite{frenje2019} found that the electronic (Brown--Preston--Singleton, BPS) stopping underpredicts the measured ion stopping near the Bragg peak in a DT plasma at $T_e = 2$~keV, and attributed the gap to nuclear-elastic scattering and possibly coupled ion modes --- the projectile-ion/target-ion channel added in this work. Fig.~\ref{fig:24} overlays the model for a proton in DT plasma on their digitized Fig.~5, with a single areal-density scale fixed by matching the model electronic stopping to BPS. With the nuclear/ion term the model reproduces the measured stopping within error at three of the four probe energies; at the lowest velocity (0.19 MeV/u) the nuclear/ion channel lifts the stopping about 12\% above the electronic value, in the direction of the measured low-velocity excess, and grows toward lower velocity. The areal-density scale is not a free parameter: digitizing their measured $T_e(r)$ and $n_e(r)$ profiles gives a path-integrated electron areal density of $(3.6$--$7.1)\times10^{21}$~cm$^{-2}$, and the scale that aligns the model with BPS, $5.5\times10^{21}$~cm$^{-2}$, lies within that range. This is the strongest experimental motivation for the nuclear/ion addition.

\section{Energy deposition and the practical sufficiency of the average-atom model}\label{hedlp-deposition}

Electronic stopping is not a resonant process. In the Fokker--Planck picture from which the stopping power derives, the energy loss is the cumulative result of a great many small-angle Coulomb collisions; what is measured and what governs transport is the integrated energy loss and range, not the excitation energy of any particular bound orbital. An average-atom dielectric model such as e-RPA, which treats all bound and free electrons self-consistently, therefore, already captures the physics that controls energy deposition. The residual differences between the average-atom and first-principles models --- the \textasciitilde10--20\% spread near the Bragg peak among e-RPA, RPA, and TD-DFT seen in Fig.~\ref{fig:1} and \ref{fig:11} --- are real, but of limited practical consequence once the finite quality of any real ion source is considered.

The depth--deposition (Bragg) profile of a 10 MeV proton beam stopping in aluminum, built from the total (electronic + Gordon--Kim nuclear) e-RPA stopping and convolved over a Gaussian energy distribution of several relative widths, makes this quantitative. The peak for a monoenergetic beam is sharp; a 1\% energy spread --- characteristic of an RF accelerator, or of the thermal Doppler broadening of fusion-product energies --- already reduces the peak deposition to 37\% of its monoenergetic value and broadens it sixteen-fold, and a 10--20\% spread, typical of laser-wakefield (LWFA) or target-normal-sheath-acceleration (TNSA) proton beams, washes the peak out almost entirely. Against this smearing, a representative 15\% stopping-model difference (the e-RPA--RPA / AA--TD-DFT spread) changes the deposited peak by 14.8\% for a monoenergetic beam but by only 0.7\% once a 10\% spread is applied. Even an ideal monoenergetic beam cannot produce an arbitrarily sharp peak: the intrinsic Bohr range straggling ($\sigma_R = 8.5$~$\mu$m, 1.4\% of the range here) sets a floor equivalent to \textasciitilde1\% energy spread.

The divergence of a finite beam acts similarly through geometry: a ray launched at angle $\theta$ to the axis reaches an axial depth reduced by $\cos\theta$ and deposits its energy at the higher rate $S/\cos\theta$, so a divergence distribution smears and shifts the Bragg peak just as an energy spread does. For 10 MeV protons in aluminum, a 10-degree rms divergence --- typical of LWFA and fast-ignition proton beams --- lowers the peak from $447$ to $179$~MeV/(g\,cm$^{-2}$); combined with a 10\% energy spread, it flattens the peak to 73, and a broad TNSA-like source (30\%, 20 degrees) to 51, while pulling it toward the surface. In the fusion context, this reaches its limit: alphas and other fast charged particles are born isotropically $(4 \pi)$, distributed through the burning volume, and with a thermal energy spread, so their deposition is fully volumetric and the Bragg peak does not exist at all; what then matters is the retained (self-heated) fraction, fixed by the alpha range relative to the hot-spot areal density $\rho R$, again an integrated quantity that the e-RPA range gives accurately.

\begin{figure*}[tp]
\centering
\includegraphics[width=\linewidth]{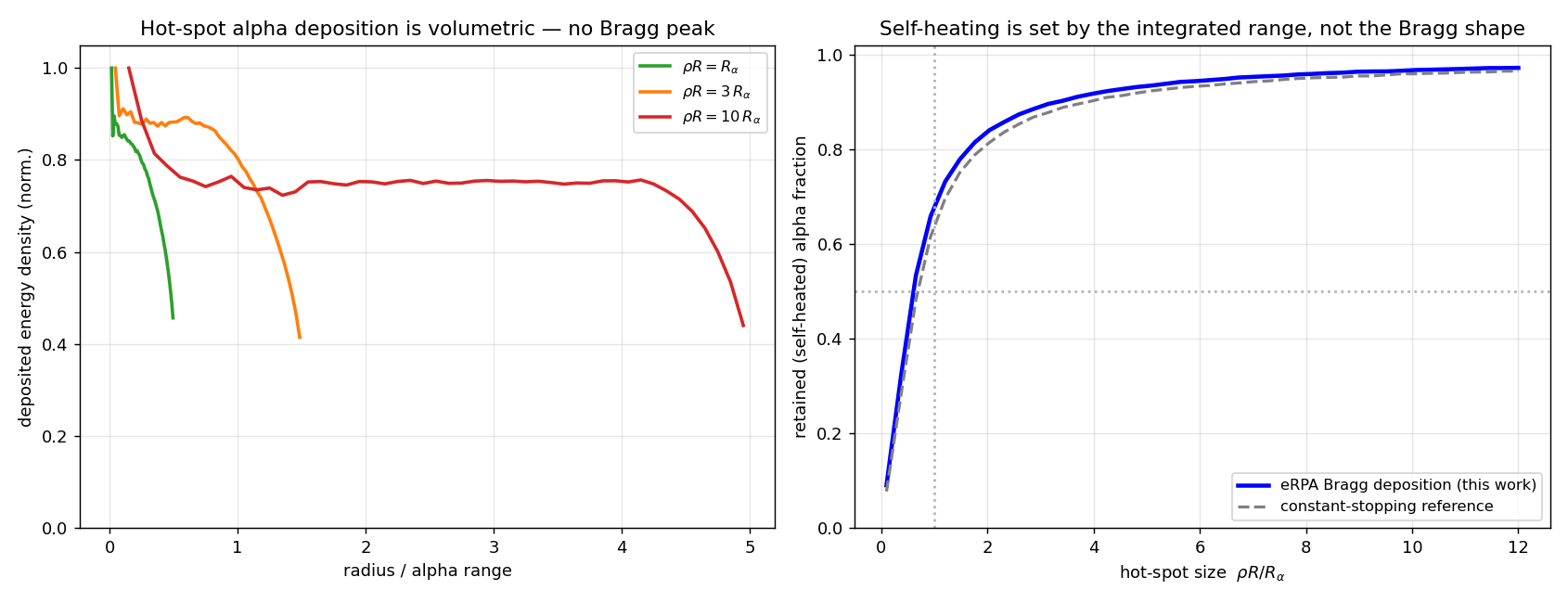}
\caption{Alpha deposition in a DT hot spot (3-D Monte Carlo; 3.5 MeV alphas, $\rho = 100$~g/cc, $T = 10$~keV). Left: deposited energy density vs radius is volumetric, with no Bragg peak. Right: retained (self-heated) alpha fraction vs hot-spot size $\rho R/R_\alpha$ (e-RPA Bragg deposition; constant-stopping reference dashed).}
\label{fig:23}
\end{figure*}

This limit is quantified in Fig.~\ref{fig:23}, a three-dimensional Monte-Carlo transport of 3.5 MeV alphas born uniformly and isotropically through a uniform DT hot spot ($\rho = 100$~g/cc, $T = 10$~keV; alpha range $R_{\alpha} = 0.083$~g/cm$^2$), using the e-RPA Bragg deposition profile. The deposited energy density is volumetric --- flat through the interior and rounded over one alpha range at the edge --- with no Bragg peak at any hot-spot size. The retained (self-heated) fraction depends only on the ratio $\rho R / R_{\alpha}$, reaching about 0.7 when $\rho R$ equals the alpha range, in agreement with the classic inertial-confinement result \cite{krokhin1973,atzeni2004}, and is insensitive to the detailed deposition profile (a constant-stopping reference agrees to within a few percent). What governs alpha self-heating is therefore the integrated alpha range --- which the e-RPA model provides accurately --- not the differential Bragg shape.

These observations also frame the relation of the present model to SRIM (Ziegler--Biersack--Littmark), the de-facto standard for cold-matter ion ranges. SRIM obtains its electronic stopping from a semi-empirical fit to the experimental database, its nuclear stopping from a universal screened-Coulomb potential, and its ranges and projected ranges from a Monte-Carlo binary-collision transport --- all restricted to cold matter. The e-RPA model is complementary: for cold targets, it reproduces the PSTAR/SRIM-family results (Fig.~\ref{fig:1} and \ref{fig:18}) from a physics-based dielectric calculation rather than a fit, and it extends the same framework continuously to the warm and hot dense plasmas, and arbitrary ionization, that SRIM does not treat. Where the two differ near the Bragg peak, the difference is within the model spread that, as shown above, is washed out by any realistic source.

\begin{figure*}[tp]
\centering
\includegraphics[width=\linewidth]{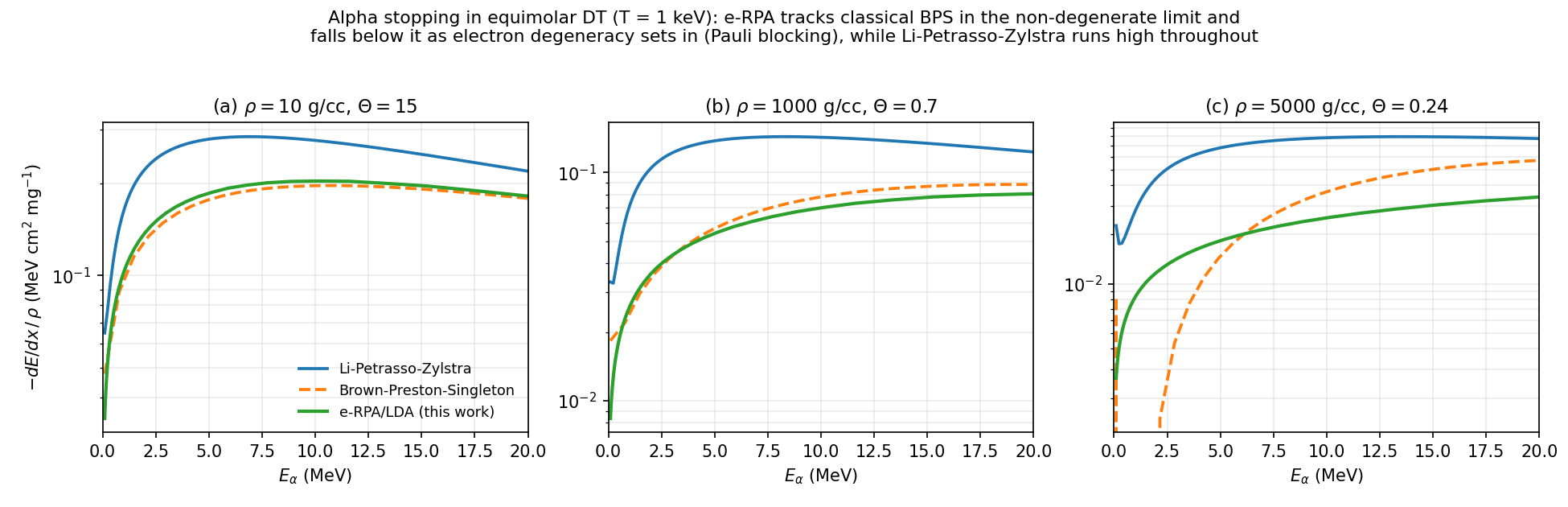}
\caption{Alpha stopping power in equimolar DT at $T=1$~keV from three models, at increasing density and electron degeneracy ($\Theta=T/E_F$): (a) non-degenerate ($\rho=10$~g/cc, $\Theta=15$), (b) degenerate ($\rho=1000$~g/cc, $\Theta=0.7$), and (c) strongly degenerate ($\rho=5000$~g/cc, $\Theta=0.24$). e-RPA/LDA (this work) coincides with the classical Brown--Preston--Singleton result when the plasma is non-degenerate and falls progressively below it as degeneracy sets in (Pauli blocking), while the Li--Petrasso--Zylstra model runs high throughout. BPS, which assumes classical statistics, also becomes unreliable at low projectile velocity in the degenerate panels. The Li--Petrasso--Zylstra and Brown--Preston--Singleton curves are computed with the implementation of Borscz et al.~\cite{Borscz2026}.}
\label{fig:degen}
\end{figure*}

The same dielectric formulation gives e-RPA a further reach that the analytic plasma-stopping models used in fusion design do not share: the high-density, degenerate regime. Because electronic stopping derives from the finite-temperature-RPA dielectric response of Maynard and Deutsch~\cite{Maynard3}, defined for an electron fluid of arbitrary degeneracy, e-RPA / LDA incorporates Fermi degeneracy and the associated Pauli blocking implicitly and self-consistently, without an effective-temperature patch; the local-density approximation carries this to the strongly degenerate densities found near the nucleus and in compressed fuel. Figure~\ref{fig:degen} illustrates the consequence for alpha stopping in equimolar DT. In the non-degenerate limit ($\rho=10$~g/cc, $\Theta\equiv T/E_F\approx15$) e-RPA/LDA coincides with the classical Brown--Preston--Singleton (BPS) result to within a few percent---an independent corroboration from two unrelated formalisms---while the binary-collision Li--Petrasso--Zylstra model runs about $40\%$ high near the peak. As the fuel becomes increasingly degenerate, however, e-RPA falls increasingly below the classical BPS curve---by about $9\%$ at $\rho=1000$~g/cc ($\Theta\approx0.7$) and $40\%$ at $\rho=5000$~g/cc ($\Theta\approx0.24$)---exactly the reduction in energy loss that electron degeneracy demands and that BPS, built on classical statistics, cannot reproduce. This is the physics Skupsky \cite{Skupsky1977} identified as essential to charged-particle energy loss in inertial-fusion fuel, where degenerate electrons reduce the stopping below its classical value and can even reflect MeV alpha particles from a Fermi-degenerate tamper, lowering the minimum ignition $\rho R$~\cite{Skupsky1980,Skupsky1977}. e-RPA/LDA thus remains valid along electron degeneracy as well, complementing its reach into partial ionization, cold matter, strong coupling, and arbitrary projectile and target charge---the regimes most relevant to advanced-fuel and high-compression fusion concepts~\cite{Borscz2026}.

Together, these results support a pragmatic conclusion: for the energy-deposition profiles and ranges that drive applications --- fusion alpha and proton heating, ion-beam fast ignition, ion-beam therapy, and radiation effects --- an accurate, efficient average-atom model such as e-RPA, combined with the nuclear/ionic term, is sufficient, because its residual differences from more expensive first-principles methods are smaller than the irreducible smearing imposed by source energy spread, divergence, and intrinsic straggling. This is not an argument against first-principles stopping calculations, which remain essential benchmarks, but a statement about where modeling effort is best spent for practical transport.

\section{Proposed experiments at the stopping-power maximum}\label{future-experiments}

The largest disagreements among stopping-power models---and the largest experimental uncertainties---occur at the stopping-power maximum, where the projectile velocity approaches the mean velocity of the target electrons. This is already true in cold matter: the curated databases show their widest scatter through the Bragg peak (gold is a notable example), reflecting the experimental difficulty of the measurement at low projectile energy. In plasmas the divergence is sharper still---perturbative and nonperturbative theories differ by $20$--$30\%$ there~\cite{Cayzac2015}---and it is precisely the regime that governs alpha-particle heating and ion-beam energy deposition.

Cayzac \textit{et al.}~\cite{Cayzac2017} demonstrated that this spread is experimentally resolvable. Probing a laser-heated carbon plasma ($n_e\approx5\times10^{20}$~cm$^{-3}$, $T_e\approx150$~eV) near the peak with $\sim1\%$ energy resolution, they disproved the standard perturbative models and found agreement only with a nonperturbative T-matrix treatment of strong ion--electron collisions. Such a measurement requires two co-diagnosed ingredients: the time-varying plasma conditions, reconstructed from radiation-hydrodynamics simulations, and the projectile charge-state distribution, which enters the stopping as $Z_b^2$ and evolves through the plasma~\cite{Rozet1996}. Recent benchmark experiments show that even the charge state is not the whole story---excited projectile configurations contribute measurably to the stopping~\cite{Zhao2021}.

It is instructive to locate the present model among the theories based on which these experiments discriminate. At the Cayzac conditions, e-RPA/LDA sits within the perturbative cluster: the plasma is weakly coupled ($\Gamma\approx0.01$) and non-degenerate ($\Theta\approx650$), so the local-field and strong-collision corrections that distinguish e-RPA from the bare random-phase approximation are inactive, leaving the model near its first-Born value. Yet, the projectile--electron coupling is strong---$\eta=Z_b e^2/\hbar v_r\approx1.5$ for fully stripped carbon at the peak---so it is the first-Born treatment in $Z_b$ that fails, exactly the regime the T-matrix is built for. In the dielectric framework, the corresponding correction is the higher-order-in-$Z_b$ (Bloch) term, and adding it to the loss function is illuminating. Applied without restriction, it diverges near the peak---over-correcting the cold Bragg peak by tens of percent---because the higher-order $Z_b$ expansion fails there as the stopping number becomes small. Regularized instead by a velocity-based (rather than medium-based) cutoff that vanishes at the cold-matter peaks, it leaves the cold C, Al, Ag, and Au stopping unchanged while reducing the higher-velocity plasma-peak stopping by about $7\%$ --roughly a quarter of the way to the T-matrix; this correction is included in the released model. The remainder---the deepest nonperturbative regime, right at the peak where the series ceases to converge---is intrinsic, which is precisely why a fully nonperturbative treatment is required there and why the peak is the decisive experimental target. The regularized correction thus recovers the leading nonperturbative-$Z_b$ physics at a small fraction of the cost of a T-matrix, while marking the boundary beyond which only measurement, or full resummation, can go. This is consistent with the average-atom overshoot near the peak noted in Sec.~\ref{wdm-validation} and with the practical-sufficiency argument of Sec.~\ref{hedlp-deposition}: the residual peak discrepancy is real but, for integrated energy deposition, smaller than the smearing imposed by any realistic source.

The next generation of measurements can extend this benchmarking across the parameter space. Beyond the heavy-ion-plus-laser configuration at GSI, laser-driven ion sources at high-repetition-rate facilities---ELI/ELIMAIA, CLPU/VEGA, and the U.S.\ warm-dense-matter program---offer compact pump-and-probe access to a broad range of conditions, supported by the isochronous energy-selecting diagnostics now under development~\cite{Volpe2024}. These platforms carry the caveat raised in Sec.~\ref{summary-and-future-work}: a laser-accelerated beam is not as cleanly isolated as a traditional accelerated beam, so careful characterization of the accelerating fields and electron transport---together with rad-hydro reconstruction of the target and a plasma charge-state model---is a prerequisite for an unambiguous stopping measurement. The e-RPA/LDA model supplies the efficient, physics-based prediction across this space; the experiments, in turn, would determine whether its peak-region corrections close the gap to the nonperturbative result and tighten the constraint at the conditions of inertial-fusion ignition.

\section{Summary and future work}\label{summary-and-future-work}

We have reported a wide-range electronic stopping power model based on the random-phase-approximation (RPA) dielectric response of Wang et al.~\cite{Wang1} and the local-density approximation (LDA) with electron-density distributions computed in an average-atom model using the Flexible Atomic Code (FAC). The cold-matter accuracy has been substantially improved by extensions to the RPA: a strong-collision (binary-collision) correction for large momentum transfer~\cite{Zwicknagel4}, a static local-field correction~\cite{Ichimaru5}, an electron-binding correction, and the higher-order Barkas and Bloch terms, the latter regularized by a velocity cutoff so it recovers the leading nonperturbative projectile-charge physics in plasmas without disturbing the cold-matter fits. We have further added a nuclear/ionic (elastic ion--ion) stopping term~\cite{Faussurier12} and a band-gap correction for cold insulators, yielding a complete, self-consistent total stopping power that is continuous from cold solids through warm dense matter to fusion plasmas. The model was validated against the NIST PSTAR and IAEA databases for elements and compounds, the charged-particle transport-workshop and TD-DFT benchmarks, and the first light-ion plasma-stopping measurements; the e-RPA corrections reduce the mean deviation from the cold databases from about 12\% (RPA) to 4\%, while the model remains computationally efficient and applicable to high-Z targets at arbitrary ionization and degeneracy.

The model is released as the open-source package \texttt{dedx-erpa}, available on GitHub at \url{https://github.com/dedx-erpa/dedx}, with tabulated proton stopping powers and ranges for cold targets in the \texttt{data/} subdirectory; the package contents, command-line interface, and output formats are documented in the Appendix~\ref{appendix-code}. We plan to submit the e-RPA/LDA package to the PlasmaPy developers as an Affiliated Package, in order to increase its visibility within the plasma-physics community.

Several extensions are planned. We have had requests to incorporate the model into a version of PRISM's Spect3D that imports plasma conditions from hydrodynamic simulations to facilitate experimental data analysis and to implement it as an efficient ion-energy-deposition module in radiation-hydrodynamics and hybrid-PIC Monte-Carlo codes---PRISM's HELIOS-CR would be a natural test bed, and both the WARP-X (LBNL/LLNL) and Triforce (UR/LLE) developers have expressed interest. The Sandia analysis of the rad-hydrodynamics relied on hydrocode reconstructions of the time-varying target (Sec.~\ref{wdm-validation}), and modern laser-plasma stopping experiments likewise derive their plasma conditions from two-dimensional radiation-hydrodynamics simulations~\cite{Cayzac2015}, so a rad hydro re-analysis of the Young and Olsen experiments---and forward-modeling of the next generation of ultrashort-pulse-laser platforms---would tighten the inferred conditions and sharpen the comparison with the model. Supporting the renewed interest in heavy-ion drivers for inertial fusion energy, which is discussed in the IFE Basic Needs Workshop Report~\cite{Ma5380}, will require an open-source model for the projectile charge state. In cold matter the charge state evolution of a fast ion is well described by codes such as ETACHA~\cite{Rozet1996}; in a plasma, where strong beam-plasma coupling drives nonequilibrium charge states and the projectile charge enters the stopping quadratically, the distribution must be evolved through the time-varying conditions, as demonstrated for laser-generated plasmas by Cayzac \textit{et al.}~\cite{Cayzac2015}. Such a model is directly analogous to the collisional-radiative models that FAC already supports, so the necessary infrastructure is in place.

The verification-and-validation chain pursued here -- cold matter $\rightarrow$ warm dense matter $\rightarrow$ hot plasma -- makes plain that the principal limitation on stopping-power modeling is now the data, not the models. We therefore close with a call for new measurements across all three regimes. In normal matter, the decline in published stopping data since 2015~\cite{Montari5969} has left unresolved conflicts between datasets and large gaps in projectile--target-energy coverage; high-quality measurements are needed both to settle those conflicts and to guard against the propagation of systematic errors that arises when semi-empirical codes are used to analyze the very experiments that calibrate them. In warm dense matter, the data remain sparsest precisely where the models diverge most---near the Bragg peak, at $v_p/v_{th}\lesssim 10$ and $\Gamma\sim 0.1$--$10$ (Fig.~\ref{fig:paramspace})---the region that governs ion-beam heating and the cold-fuel leg of fusion deposition, and the region the historic NRL and Sandia data already begin to populate. And in the high-energy-density regime, the achievement of ignition~\cite{Abu-6018} now provides a highly diagnosed platform for constraining the temperature dependence of alpha stopping in dense DT, complementing the dedicated low-Z plasma-stopping experiments underway at facilities in the US, Europe, and China.

The emerging generation of high-repetition-rate laser and pulsed-power facilities is well suited to this task: by accumulating many shots, they can build the statistics needed to map stopping power across temperature, density, and ionization and to fill the underused low-$v_p/v_{th}$ region of the parameter space. This advantage is greatest for normal matter, where the target is static and need not be replaced or regenerated between shots: a high-repetition-rate beam can accumulate large numbers of measurements on a single sample under identical conditions, directly reducing the database scatter that is widest at the Bragg peak (Sec.~\ref{cold-validation})---whereas for a plasma target each shot must recreate and re-diagnose the transient state, so repetition rate buys reproducibility and averaging rather than free statistics. However, we add caution, drawn from the same light-ion-fusion experience that motivates this work. Ultrashort-pulse-laser ion sources are not as cleanly isolated as a traditional accelerated beam: the accelerating sheath fields and the accompanying energetic electrons can co-propagate with---and outrun---the ions, so that the target may be heated, pre-ionized, or otherwise perturbed by agents other than the ion beam whose stopping is being measured. Extracting an unambiguous stopping power on such platforms therefore requires careful characterization of these fields and of the electron transport, or experimental geometries that separate the diagnostic ion from its own source. Traditional accelerators and pulsed-power-driven ion beams, which deliver a comparatively clean and well-isolated probe, retain a distinct and complementary value for this purpose. Together with the open-source release of our model, we hope this work encourages both new experiments in these regimes and a quantitative re-examination of the existing data.

\section{Acknowledgements}\label{acknowledgements}

This work was partially supported by the U.S. Department of Energy, Office of Science, Fusion Energy Sciences (FES) under Award Number DE-SC0022112. We thank Sophia Malko, Luca Volpe, Johan Frenje, and Will Fox for discussions about stopping power experiments. The authors also thank Suxing Hu for discussions about DFT and modifications to AA for compound materials and Marcus Borscz for sharing comparisons with modified Lee More and BPS. The authors acknowledge the use of Anthropic's Claude as a coding, analysis, and manuscript-preparation assistant under the authors' direction; the authors verified all outputs and take full responsibility for the content.

\newcommand{\opt}[1]{\texttt{-{}-#1}}
\appendix

\section{Open-source code: contents and usage}\label{appendix-code}

The e-RPA/LDA electronic model, together with the nuclear/ionic and cold-insulator band-gap extensions, is implemented in the open-source package \texttt{dedx-erpa}, available at \url{https://github.com/dedx-erpa/dedx}. This appendix reproduces the package documentation: contents, installation, command-line interface, and output formats. The package is free software, distributed under the GNU General Public License, version 3 (GPLv3); the full license text is in the repository's \texttt{LICENSE} file. Users are free to run, study, modify, and redistribute the code under the terms of that license, which requires that derivative works also be released under the GPL. Because the package builds on the Flexible Atomic Code, itself released under GPLv3, it adopts the same license for consistency and compatibility.

\subsection{Package contents}

\begin{description}
\item[\texttt{dief.py}] computes the RPA dielectric functions and the stopping power of ions in a uniform electron gas, used to tabulate the stopping numbers on a temperature--density grid (\texttt{t\#\#.dat}).
\item[\texttt{dedx.f}] a Fortran program that convolves the e-RPA stopping numbers with the electron-density distribution to produce the stopping power of any atomic species.
\item[\texttt{dedx.py}] the driver: computes the average-atom electron-density distribution and invokes \texttt{dedx.f} to generate the stopping powers.
\item[\texttt{nuclear.py}, \texttt{nuclear\_gk.py}] the nuclear/ionic stopping (elastic ion--ion at low temperature; projectile-ion/target-ion Fokker--Planck drag at finite temperature; Faussurier \textit{et al.}~\cite{Faussurier12}), which adds to the electronic term to give a total stopping power.
\item[\texttt{data/}] proton stopping in cold targets for $Z=1$--$92$ and selected compounds; each material has a sub-directory (named by chemical symbol) holding \texttt{dedx.dat} and plots, with independent reference data in \texttt{data/refs/}.
\end{description}

\subsection{Installation}

The package requires \texttt{numpy}, \texttt{scipy}$\,\ge\,$1.14, and \texttt{matplotlib} (see \texttt{requirements.txt}), plus \texttt{pfac}, the Python interface to FAC, installed from the master branch of \url{https://github.com/flexible-atomic-code} (earlier FAC releases are not compatible). After installing FAC, edit the \texttt{Makefile} and compile \texttt{dedx.f} with \texttt{make}.

\subsection{Command-line interface}

The principal options of \texttt{dedx.py} are
\begin{description}
\item[\opt{zp}] projectile atomic number (default 1).
\item[\opt{zt}] target atomic number.
\item[\opt{zc}, \opt{wc}] for compounds, comma-separated component atomic numbers and their number-weights.
\item[\opt{fc}] alternatively, the chemical formula of the compound, e.g.\ \texttt{Al2O3}.
\item[\opt{d}, \opt{t}] target mass density (g/cc) and temperature (eV).
\item[\opt{od}] output directory.
\item[\opt{emin}, \opt{emax}, \opt{mep}] projectile-energy range (MeV) and number of points.
\item[\opt{aa}] average-atom mode: 2 runs the AA model to generate the density distribution; 1 reuses a previous AA run; 0 reads an existing density file.
\item[\opt{mloss}] stopping-power mode. Mode 0 uses the analytic RPA fit of Wang et al.; modes 1--4 select bare RPA, $+$local-field, $+$strong-collision, and $+$both corrections; adding 10 includes the Barkas term and 20 the Barkas \emph{and} Bloch terms (the latter regularized by a velocity cutoff so it vanishes at the cold-matter peaks). The default, \texttt{mloss=24} (local-field $+$ strong-collision $+$ Barkas $+$ Bloch, with the bound-electron correction), is the recommended mode.
\item[\opt{mout}] if 1, also output the radius-resolved contribution to $dE/dx$.
\end{description}

\subsection{Output}

A run writes \texttt{dedx.dat} to the output directory. A header (lines beginning with \texttt{\#}) records the target composition, projectile charge, unit-cell radius $r_s$, electron temperature, density, and mean ionization $\bar{Z}$; the data section has three columns: energy per nucleon (MeV), electronic stopping power ($10^{-15}$~eV~cm$^2$/atom), and range (mg/cm$^2$).

\subsection{Nuclear/ionic stopping}

Adding \opt{nuc}=1 computes the nuclear/ionic stopping and writes \texttt{dedx\_nuc.dat}. The potential is selected with \opt{npot} (subset of \texttt{gk}, \texttt{ionsphere}, \texttt{yukawa}; default \texttt{gk}): the full Gordon--Kim model \texttt{gk} builds the projectile--target interaction from the average-atom bound-plus-free electron density with the complete electron-gas energy and reproduces the cold NIST nuclear stopping, whereas the analytic \texttt{ionsphere}/\texttt{yukawa} potentials screen with free electrons only. The ion temperature is set with \opt{ti} (default $=T_e$); the finite-$T_i$ Maxwellian average captures the thermal motion of the target ions and the energy-gain regime below the thermalization threshold. The output adds per-potential nuclear columns, the total stopping, and both the CSDA pathlength and the projected (practical) range obtained from the nuclear momentum-transfer cross section.

\subsection{Examples and citation}

Representative runs:
\begin{description}
\item[Cold Al] \texttt{python dedx.py \opt{zt}=13 \opt{aa}=2 \opt{d}=2.7 \opt{t}=0.025 \opt{od}=ColdAl}
\item[Warm Mylar] \texttt{python dedx.py \opt{fc}=H4C5O2 \opt{aa}=2 \opt{d}=1.35 \opt{t}=10.0 \opt{od}=MylarWDM}
\item[Cold Al, total] \texttt{python dedx.py \opt{zt}=13 \opt{d}=2.7 \opt{t}=0.025 \opt{aa}=2 \opt{nuc}=1 \opt{npot}=gk \opt{od}=ColdAl}
\end{description}
The repository also provides validation drivers (\texttt{nuc\_test.py}, \texttt{nuc\_fac\_compare.py}, \texttt{alpha\_dt\_verify.py}, and others) that reproduce the cold-NIST nuclear comparison, the warm-dense-matter temperature sweep, and the alpha-in-DT crossover. When using the package, please cite the e-RPA electronic model of Wang, Mehlhorn, and MacFarlane~\cite{Wang1}, the present work for the corrections and the cold-insulator band-gap treatment, and Faussurier \textit{et al.}~\cite{Faussurier12} for the nuclear/ionic stopping.

\bibliographystyle{unsrt}

\bibliography{references.bib}

\begin{thebibliography}{10}

\bibitem{Berger6130}
M.~J. Berger, J.S. Coursey, M.A. Zucker, and J.~Chang.
\newblock (2005), estar, pstar, and astar: Computer programs for calculating
  stopping-power and range tables for electrons, protons, and helium ions
  (version 1.2.3), June 17, 2026 2005.

\bibitem{Montari5969}
C.~C. Montanari, P.~Dimitriou, L.~Marian, A.~M.~P. Mendez, J.~P. Peralta, and
  F.~Bivort-Haiek.
\newblock The iaea electronic stopping power database: Modernization, review,
  and analysis of the existing experimental data.
\newblock {\em Nuclear Instruments and Methods in Physics Research Section B:
  Beam Interactions with Materials and Atoms}, 551:165336, 2024.

\bibitem{Ziegler6266}
J.~F. Ziegler, M.D. Ziegler, and U.~Littmark.
\newblock {\em SRIM - The Stopping and Range of Ions in Matter}, volume 1 of
  The Stopping and Range of ions in Matter.
\newblock Pergamon Press, 2013.

\bibitem{ICRU6267}
International Committee for~Radiological Units.
\newblock Icru report 49 - stopping power and ranges for protons and alpha
  particles.
\newblock
  \url{https://nds.iaea.org/stopping-legacy/stopping_201812/stopping_heli.html}.

\bibitem{Anderson4330}
Hans.H. Anderson and James.F. Ziegler.
\newblock {\em HYDROGEN: The Stopping and Ranges of Ions in Matter}.
\newblock Pergamon, 1977.

\bibitem{Fraley1287}
G.~S. Fraley, E.~J. Linnebur, R.~J. Mason, and R.~L. Morse.
\newblock Thermonuclear burn characteristics of compressed deuterium-tritium
  microspheres.
\newblock {\em Physics of Fluids}, 17(2):474--489, 1974.

\bibitem{Zylstra6268}
A.~B. Zylstra and O.~A. Hurricane.
\newblock On alpha-particle transport in inertial fusion.
\newblock {\em Physics of Plasmas}, 26(6):062701, 2019.

\bibitem{Ma5380}
T.~Ma and R.~Betti.
\newblock Inertial fusion energy: Report of the 2022 fusion energy sciences
  basic research needs workshop, June 21st -- 23rd, 2022 2022.

\bibitem{Casas8}
D.~Casas, M.~D. Barriga-Carrasco, and J.~Rubio.
\newblock Evaluation of slowing down of proton and deuteron beams in {CH2},
  {LiH}, and {Al} partially ionized plasmas.
\newblock {\em Physical Review E}, 88(3):033102, 2013.

\bibitem{Malko9}
S.~Malko, W.~Cayzac, V.~Ospina-Boh{\'o}rquez, et~al.
\newblock Proton stopping measurements at low velocity in warm dense carbon.
\newblock {\em Nature Communications}, 13(1):2893, 2022.

\bibitem{Reichelt5864}
Benjamin~L. Reichelt, Richard~D. Petrasso, and Chikang Li.
\newblock Effects of alpha-ion stopping on ignition and ignition criteria in
  inertial confinement fusion experiments.
\newblock {\em Physics of Plasmas}, 31(1), 2024.

\bibitem{Barriga5439}
M.~D. Barriga-Carrasco, J.~Vázquez-Moyano, and F.~Chacón-Rubio.
\newblock Calculations on the stopping power of the warm dense matter at the
  bragg peak.
\newblock {\em Physics Letters A}, 447:128294, 2022.

\bibitem{Young10}
F.~C. Young, D.~Mosher, S.~J. Stephanakis, S.~A. Goldstein, and T.~A. Mehlhorn.
\newblock Measurements of enhanced stopping of 1-{MeV} deuterons in
  target-ablation plasmas.
\newblock {\em Physical Review Letters}, 49(8):549--553, 1982.

\bibitem{Olsen11}
J.~N. Olsen, T.~A. Mehlhorn, J.~Maenchen, and D.~J. Johnson.
\newblock Enhanced ion stopping powers in high-temperature targets.
\newblock {\em Journal of Applied Physics}, 58(8):2958--2967, 1985.

\bibitem{atzeni2004}
S.~Atzeni and J.~Meyer-ter Vehn.
\newblock {\em The Physics of Inertial Fusion}.
\newblock Oxford University Press, 2004.

\bibitem{Ren4167}
Jieru Ren, Zhigang Deng, Wei Qi, Benzheng Chen, Bubo Ma, Xing Wang, Shuai Yin,
  Jianhua Feng, Wei Liu, Zhongfeng Xu, Dieter H.~H. Hoffmann, Shaoyi Wang,
  Quanping Fan, Bo~Cui, Shukai He, Zhurong Cao, Zongqing Zhao, Leifeng Cao,
  Yuqiu Gu, Shaoping Zhu, Rui Cheng, Xianming Zhou, Guoqing Xiao, Hongwei Zhao,
  Yihang Zhang, Zhe Zhang, Yutong Li, Dong Wu, Weimin Zhou, and Yongtao Zhao.
\newblock Observation of a high degree of stopping for laser-accelerated
  intense proton beams in dense ionized matter.
\newblock {\em Nature Communications}, 11(1):5157, 2020.

\bibitem{frenje2019}
J.~A. Frenje, P.~E. Grabowski, C.~K. Li, et~al.
\newblock Measurements of ion stopping around the {Bragg} peak in
  high-energy-density plasmas.
\newblock {\em Physical Review Letters}, 122(1):015002, 2019.

\bibitem{Abu-6018}
H.~Abu-Shawareb, R.~Acree, P.~Adams, J.~Adams, B.~Addis, R.~Aden, P.~Adrian,
  B.~B Afeyan, M.~Aggleton, L.~Aghaian, A.~Aguirre, D.~Aikens, J.~Akre,
  F.~Albert, M.~Albrecht, B.~J Albright, J.~Albritton, J.~Alcala, C.~Alday,
  D.~A Alessi, N.~Alexander, J.~Alfonso, N.~Alfonso, E.~Alger, S.~J Ali, Z.~A
  Ali, A.~Allen, W.~E Alley, P.~Amala, P.~A Amendt, P.~Amick, S.~Ammula,
  C.~Amorin, D.~J Ampleford, R.~W Anderson, T.~Anklam, N.~Antipa, B.~Appelbe,
  C.~Aracne-Ruddle, E.~Araya, T.~N Archuleta, M.~Arend, P.~Arnold, T.~Arnold,
  A.~Arsenlis, J.~Asay, L.~J Atherton, D.~Atkinson, R.~Atkinson, J.~M Auerbach,
  B.~Austin, L.~Auyang, A.~A~S Awwal, N.~Aybar, J.~Ayers, S.~Ayers, T.~Ayers,
  S.~Azevedo, B.~Bachmann, C.~A Back, J.~Bae, D.~S Bailey, J.~Bailey,
  T.~Baisden, K.~L Baker, H.~Baldis, D.~Barber, M.~Barberis, D.~Barker,
  A.~Barnes, C.~W Barnes, M.~A Barrios, C.~Barty, I.~Bass, S.~H Batha, S.~H
  Baxamusa, G.~Bazan, J.~K Beagle, R.~Beale, B.~R Beck, J.~B Beck, M.~Bedzyk,
  R.~G Beeler, R.~G Beeler, W.~Behrendt, L.~Belk, P.~Bell, M.~Belyaev, J.~F
  Benage, G.~Bennett, L.~R Benedetti, L.~X Benedict, R.~L Berger, T.~Bernat,
  L.~A Bernstein, B.~Berry, L.~Bertolini, G.~Besenbruch, J.~Betcher,
  R.~Bettenhausen, et~al.
\newblock Achievement of target gain larger than unity in an inertial fusion
  experiment.
\newblock {\em Physical Review Letters}, 132(6), 2024.

\bibitem{Grab4212}
P.~E. Grabowski, S.~B. Hansen, M.~S. Murillo, L.~G. Stanton, F.~R. Graziani,
  A.~B. Zylstra, S.~D. Baalrud, P.~Arnault, A.~D. Baczewski, L.~X. Benedict,
  C.~Blancard, O.~Certik, J.~Clerouin, L.~A. Collins, S.~Copeland, A.~A.
  Correa, J.~Dai, J.~Daligault, M.~P. Desjarlais, M.~W.~C. Dharma-wardana,
  G.~Faussurier, J.~Haack, T.~Haxhimali, A.~Hayes-Sterbenz, Y.~Hou, S.~X. Hu,
  D.~Jensen, G.~Jungman, G.~Kagan, D.~Kang, J.~D. Kress, Q.~Ma, M.~Marciante,
  E.~Meyer, R.~E. Rudd, D.~Saumon, L.~Shulenburger, R.~Singleton, T.~Sjostrom,
  L.~J. Stanek, C.~E. Starrett, C.~Ticknor, S.~Valaitis, J.~Venzke, and
  A.~White.
\newblock Review of the first charged-particle transport coefficient comparison
  workshop.
\newblock {\em High Energy Density Physics}, 37:29, 2020.

\bibitem{Stanek13}
L.~G. Stanek et~al.
\newblock Review of the second charged-particle transport coefficient code
  comparison workshop.
\newblock {\em Physics of Plasmas}, 31, 2024.

\bibitem{Mehl459}
T.~A. Mehlhorn.
\newblock Intense ion beams for inertial confinement fusion.
\newblock {\em IEEE Transactions on Plasma Science}, 25(6):1336--56, 1997.

\bibitem{Quintenz1514}
J.~P. Quintenz, J.~E. Bailey, K.~W. Bieg, D.~D. Bloomquist, D.~L. Cook, J.~T.
  Crow, R.~S. Coats, M.~S. Derzon, M.~P. Desjarlais, P.~L. Dreike, R.~A.
  Gerber, T.~W. Hussey, D.~J. Johnson, W.~A. Johnson, R.~P. Kensek, G.~W.
  Kuswa, J.~R. Lee, R.~J. Leeper, T.~R. Lockner, J.~E. Maenchen, D.~H.
  McDaniel, P.~F. McKay, T.~A. Mehlhorn, Jr. Mendel, C.~W., L.~P. Mix, E.~L.
  Neau, C.~L. Olson, T.~D. Pointon, A.~L. Pregenzer, T.~J. Renk, G.~E. Rochau,
  S.~E. Rosenthal, C.~L. Ruiz, L.~X. Schneider, S.~A. Slutz, R.~W. Stinnett,
  W.~A. Stygar, M.~A. Sweeney, F.~C. Tisone, B.~N. Turman, J.~P. VanDevender,
  and J.~R. Woodworth.
\newblock Status of the pbfa-ii light ion beam fusion program.
\newblock In {\em Particle Accelerator Conference, 1989. Accelerator Science
  and Technology., Proceedings of the 1989 IEEE}, pages 752--756 vol.2, 1989.

\bibitem{Mehlhorn447}
T.~A. Mehlhorn.
\newblock A finite material temperature model for ion energy deposition in
  ion-driven inertial confinement fusion targets.
\newblock {\em Journal of Applied Physics}, 52(11):6522--32, 1981.

\bibitem{Mehlhorn1147}
T.~A. Mehlhorn, J.~M. Peek, E.~J. McGuire, J.~N. Olsen, and F.~C. Young.
\newblock Current status of calculations and measurements of ion stopping power
  in icf plasmas.
\newblock {\em Journal De Physique}, 44(NC-8):39--66, 1983.

\bibitem{Maenchen3515}
J.~Maenchen, T.~A. Mehlhorn, D.~F. Wenger, R.~J. Leeper, D.~J. Johnson, and
  T.~R. Lockner.
\newblock Intense ion beam kα measurements on pbfa‐ii.
\newblock {\em Review of Scientific Instruments}, 59(8):1706--1708, 1988.

\bibitem{Bailey4030}
J.~Bailey, A.~L. Carlson, G.~Chandler, M.~S. Derzon, R.~J. Dukart, B.~A.
  Hammel, D.~J. Johnson, T.~R. Lockner, J.~Maenchen, E.~J. McGuire, T.~A.
  Mehlhorn, W.~E. Nelson, L.~E. Ruggles, W.~A. Stygar, and D.~F. Wenger.
\newblock Observation of kα. x-ray satellites from a target heated by an
  intense ion beam.
\newblock {\em Laser and Particle Beams}, 8(4):555--562, 1990.

\bibitem{Chandler1517}
G.~A. Chandler, J.~Aubert, J.~Bailey, A.~Carlson, D.~Derzon, M.~Derzon,
  R.~Dukart, R.~Humphreys, J.~Hunter, D.~J. Johnson, M.~K. Matzen, A.~Moats,
  R.~Olson, J.~Pantuso, P.~Rockett, C.~Ruiz, P.~Sawyer, J.~Torres, and
  T.~Hussey.
\newblock Icf target diagnostics on pbfa ii (invited).
\newblock {\em Review of Scientific Instruments}, 63(10):4828--4833, 1992.

\bibitem{Mehlh1518}
T.~A. Mehlhorn, J.~E. Bailey, G.~A. Chandler, R.~S. Coats, M.~E. Cuneo, M.~S.
  Derzon, M.~P. Desjarlais, R.~J. Dukart, A.~B. Filuk, T.~A. Haill, H.~C. Ives,
  D.~J. Johnson, R.~J. Leeper, T.~R. Lockner, C.~W. Mendel, P.~R. Menge, L.~P.
  Mix, A.~R. Moats, W.~B. Moore, T.~D. Pointon, J.~W. Poukey, J.~P. Quintenz,
  S.~E. Rosenthal, D.~Rovang, C.~L. Ruiz, S.~A. Slutz, W.~A. Stygar, and D.~F.
  Wenger.
\newblock Progress in lithium beam power, divergence, and intensity at sandia
  national laboratories.
\newblock In {\em High-Power Particle Beams, 1994 10th International Conference
  on}, volume~1, pages 53--56, 1994.

\bibitem{Derzon457}
M.~S. Derzon, G.~A. Chandler, R.~J. Dukart, D.~J. Johnson, R.~J. Leeper, M.~K.
  Matzen, E.~J. McGuire, T.~A. Mehlhorn, A.~R. Moats, R.~E. Olson, and C.~L.
  Ruiz.
\newblock Li-beam-heated hohlraum experiments at particle beam fusion
  accelerator ii.
\newblock {\em Physical Review Letters}, 76(3):435--8, 1996.

\bibitem{MacFarlane353}
J.J. MacFarlane, P.~Wang, J.~Bailey, T.A. Mehlhorn, R.J. Dukart, and R.C.
  Mancini.
\newblock Analysis of ka line emission from aluminum plasmas created by intense
  proton beams.
\newblock In {\em Physical Review E}, volume 47, (4), pages 2748--2758. The
  American Physical Society, 1993.

\bibitem{McGuire1148}
E.~J. McGuire.
\newblock The proton stopping power of aluminum and nickel ions.
\newblock {\em Journal of Applied Physics}, 70(12):7213--7216, 1991.

\bibitem{Wang1}
P.~Wang, T.~A. Mehlhorn, and J.~J. MacFarlane.
\newblock A unified self-consistent model for calculating ion stopping power in
  {ICF} plasma.
\newblock {\em Physics of Plasmas}, 5(8):2977--2987, 1998.

\bibitem{Liberman3892}
D.~A. Liberman, D.~T. Cromer, and J.~T. Waber.
\newblock Relativistic self-consistent field program for atoms and ions.
\newblock {\em Computer Physics Communications}, 2(2):107--113, 1971.

\bibitem{Maynard3}
G.~Maynard and C.~Deutsch.
\newblock Born random phase approximation for ion stopping in an arbitrarily
  degenerate electron fluid.
\newblock {\em Journal de Physique}, 46(7):1113--1122, 1985.

\bibitem{Gu2}
M.~F. Gu.
\newblock The flexible atomic code.
\newblock {\em Canadian Journal of Physics}, 86(5):675--689, 2008.

\bibitem{Zwicknagel4}
G.~Zwicknagel, C.~Toepffer, and P.-G. Reinhard.
\newblock Stopping of heavy ions in plasmas at strong coupling.
\newblock {\em Physics Reports}, 309(3):117--208, 1999.

\bibitem{Ichimaru5}
S.~Ichimaru and K.~Utsumi.
\newblock Analytic expression for the dielectric screening function of strongly
  coupled electron liquids at metallic and lower densities.
\newblock {\em Physical Review B}, 24(12):7385--7388, 1981.

\bibitem{Dejarlais1142}
M.~P. Desjarlais, J.~D. Kress, and L.~A. Collins.
\newblock Electrical conductivity for warm, dense aluminum plasmas and liquids.
\newblock {\em Physical Review E}, 66(2), 2002.
\newblock Part 2.

\bibitem{Desjarlais2583}
M.~P. Desjarlais.
\newblock Density-functional calculations of the liquid deuterium hugoniot,
  reshock, and reverberation timing.
\newblock {\em Physical Review B}, 68(6):8, 2003.

\bibitem{Desj2102}
M.~P. Desjarlais.
\newblock Density functional calculations of the reflectivity of shocked xenon
  with ionization based gap corrections.
\newblock {\em Contributions to Plasma Physics}, 45(3-4):300--304, 2005.

\bibitem{White7}
A.~J. White, L.~A. Collins, K.~Nichols, and S.~X. Hu.
\newblock Mixed stochastic-deterministic time-dependent density functional
  theory: application to stopping power of warm dense carbon.
\newblock {\em Journal of Physics: Condensed Matter}, 34(17):174001, 2022.

\bibitem{Esbensen6}
H.~Esbensen and P.~Sigmund.
\newblock Barkas effect in a dense medium: Stopping power and wake field.
\newblock {\em Annals of Physics}, 201(1):152--192, 1990.

\bibitem{levinelouie1982}
Z.~H. Levine and S.~G. Louie.
\newblock New model dielectric function and exchange-correlation potential for
  semiconductors and insulators.
\newblock {\em Physical Review B}, 25(10):6310--6316, 1982.

\bibitem{zylstra2015}
A.~B. Zylstra, J.~A. Frenje, P.~E. Grabowski, et~al.
\newblock Measurement of charged-particle stopping in warm dense plasma.
\newblock {\em Physical Review Letters}, 114(21):215002, 2015.

\bibitem{Faussurier12}
G.~Faussurier, C.~Blancard, and M.~Gauthier.
\newblock Nuclear stopping power in warm and hot dense matter.
\newblock {\em Physics of Plasmas}, 20(1):012705, 2013.

\bibitem{krokhin1973}
O.~N. Krokhin and V.~B. Rozanov.
\newblock Escape of $\alpha$ particles from a laser-pulse-initiated
  thermonuclear reaction.
\newblock {\em Soviet Journal of Quantum Electronics}, 2(4):393--394, 1973.

\bibitem{Skupsky1977}
Stanley Skupsky.
\newblock Energy loss of ions moving through high-density matter.
\newblock {\em Physical Review A}, 16(2):727--731, 1977.
\newblock PRA.

\bibitem{Skupsky1980}
Stanley Skupsky.
\newblock High-density effects on thermonuclear ignition for inertially
  confined fusion.
\newblock {\em Physical Review Letters}, 44(26):1760--1763, 1980.
\newblock PRL.

\bibitem{Borscz2026}
Marcus Borscz, Thomas~A. Mehlhorn, Patrick~A. Burr, Igor Morozov, and Sergey
  Pikuz.
\newblock Monte carlo simulations of suprathermal enhancement in advanced
  nuclear fusion fuels.
\newblock {\em arXiv e-prints}, page arXiv:2604.06769, 2026.

\bibitem{Cayzac2015}
W.~Cayzac, V.~Bagnoud, M.~M. Basko, A.~Bla{\v{z}}evi{\'c}, A.~Frank, D.~O.
  Gericke, L.~Hallo, G.~Malka, A.~Ortner, An. Tauschwitz, J.~Vorberger, and
  M.~Roth.
\newblock Predictions for the energy loss of light ions in laser-generated
  plasmas at low and medium velocities.
\newblock {\em Physical Review E}, 92(5):053109, 2015.

\bibitem{Cayzac2017}
W.~Cayzac, A.~Frank, A.~Ortner, V.~Bagnoud, M.~M. Basko, S.~Bedacht,
  C.~Bl{\"a}ser, A.~Bla{\v{z}}evi{\'c}, S.~Busold, O.~Deppert, J.~Ding,
  M.~Ehret, P.~Fiala, S.~Frydrych, D.~O. Gericke, L.~Hallo, J.~Helfrich,
  D.~Jahn, E.~Kjartansson, A.~Knetsch, D.~Kraus, G.~Malka, N.~W. Neumann,
  K.~P{\'e}pitone, D.~Pepler, S.~Sander, G.~Schaumann, T.~Schlegel,
  N.~Schroeter, D.~Schumacher, M.~Seibert, An. Tauschwitz, J.~Vorberger,
  F.~Wagner, S.~Weih, Y.~Zobus, and M.~Roth.
\newblock Experimental discrimination of ion stopping models near the bragg
  peak in highly ionized matter.
\newblock {\em Nature Communications}, 8:15693, 2017.

\bibitem{Rozet1996}
J.~P. Rozet, C.~Stéphan, and D.~Vernhet.
\newblock Etacha: a program for calculating charge states at ganil energies.
\newblock {\em Nuclear Instruments and Methods in Physics Research B},
  107(1-4):67--70, 1996.

\bibitem{Zhao2021}
Y.~T. Zhao, Y.~N. Zhang, R.~Cheng, B.~He, C.~L. Liu, X.~M. Zhou, Y.~Lei, Y.~Y.
  Wang, J.~R. Ren, X.~Wang, Y.~H. Chen, G.~Q. Xiao, S.~M. Savin, R.~Gavrilin,
  A.~A. Golubev, and D.~H.~H. Hoffmann.
\newblock Benchmark experiment to prove the role of projectile excited states
  upon the ion stopping in plasmas.
\newblock {\em Physical Review Letters}, 126(11):115001, 2021.

\bibitem{Volpe2024}
L.~Volpe, T.~Cebriano~Ram{\'i}rez, C.~S{\'a}nchez~S{\'a}nchez, A.~Perez,
  A.~Curcio, D.~De~Luis, G.~Gatti, B.~Kebladj, S.~Khetari, J.~A.
  Perez-Hernandez, and M.~D. Rodriguez~Frias.
\newblock A platform for ultra-fast proton probing of matter in extreme
  conditions.
\newblock {\em Sensors}, 24(16):5254, 2024.

\end{thebibliography}

\end{document}